\documentclass{aa}  
\usepackage{graphicx}
\usepackage{txfonts,textcomp}
\makeatletter
 \def\hlinewd#1{%
   \noalign{\ifnum0=`}\fi\hrule \@height #1 \futurelet
    \reserved@a\@xhline}
\makeatother

\newcommand{\kms}{km\,s$^{-1}$}
\begin{document}

    \title{Multiwavelength continuum, emission line and BAL variability with prominent $\ion{P}{v}$ 
    absorption in the X-ray-weakest quasar PG\,0043+039
    \thanks{ Based on observations obtained with XMM-Newton \& NuSTAR, HST, and the Hobby-Eberly Telescope (HET).}}

   \author{W. Kollatschny \inst{1}, 
           N. Schartel \inst{2},
           M. A. Probst \inst{1},
           M. W. Ochmann \inst{1}, 
           L. Ballo \inst{3}, 
           M. Santos-Lle\'{o} \inst{2}
           }   

   \institute{Institut f\"ur Astrophysik und Geophysik, Universit\"at G\"ottingen,
              Friedrich-Hund Platz 1, D-37077 G\"ottingen, Germany\\
              \email{wkollat@gwdg.de}
        \and
         XMM-Newton Science Operations Centre, ESA, Villafranca del Castillo, Apartado 78, 28691 Villanueva de la Cañada, Spain
        \and
        TPZ for ESA/ESAC, 28691 Villanueva de la Canada, Madrid, Spain
}

   \date{Received 1 April 2025; Accepted 3 September 2025}
   \authorrunning{Kollatschny et al.}
   \titlerunning{PG\,0043+039}

% \abstract{}{}{}{}{} 
% 5 {} token are mandatory
 
  \abstract
   { PG\,0043+039 has been identified as an extremely X-ray-weak quasar and as a peculiar broad-absorption-line quasar (BALQ) based on UV HST spectra.}
   {This study aims to investigate the spectral properties and their variation in PG\,0043+039 in more detail.}
   {We took simultaneous deep X-ray observations of PG\,0043+039 with XMM-Newton \& NuSTAR, UV spectra with the HST, and optical spectra with the HET telescope in 2022.
    }  
   {PG\,0043+039 was an extreme low-X-ray-luminosity quasar in 2022.
  This AGN showed an extreme steep $\alpha_{ox}$ value of $-$2.47. These values are similar to those measured in 2005. The X-ray luminosity was a factor of 3.4 higher in the meantime in 2013.
The optical and UV continuum only decreased by a factor of 1.3 -- 1.5 from 2013 to 2022.
Very strong emission-line intensity variations by factors of eight or more were observed in the $\ion{O}{vi}\,\lambda 1038$ and $\ion{Si}{iv}\,\lambda 1403$ lines between 2013 and 2022. The other UV emission lines such as Ly$\alpha$ only decreased by a factor of 1.4. The optical emission lines showed only minor intensity variations.
We derive black hole masses of $M=6.95\times 10^{8} M_{\odot}$ (based on H$\beta$) and 
  $M=3.53\times 10^{8} M_{\odot}$ (based on \ion{Mg}{ii}\,$\lambda2800$). This corresponds to Eddington ratios of $L/L_\text{edd}$ = 0.115 (H$\beta$) and 0.157 (\ion{Mg}{ii}\,$\lambda2800$). PG\,0043+039 exhibits strong and broad absorption lines in the UV. 
 The highly ionized absorption lines show the largest velocity blueshifts in their broad absorption lines (BALs). Lower ionized BALs have smaller blueshifts. 
The $\ion{P}{v}$ absorption is very strong, with equivalent widths of 7 -- 10 \AA{}. 
PG\,0043+039 shows strong Ly$\alpha$ emission despite strong $\ion{P}{v}$ absorption. PG\,0043+039 is the only source with strong Ly$\alpha$ in the sample of strong $\ion{P}{v}$ absorbers. 
All the strong absorption-line troughs in the UV varied in unison in velocity space back and forward 
(-2000.\,$\pm{}$300.\,\kms, +2740.\,$\pm{}$200.\,\kms) without any major changes in absorption strength or in their profiles for the years 1991, 2013, and 2022.
We found no general correlations of X-ray/opt/UV continuum variations with the broad absorption line variations. However, based
on the simultaneous UV and X-ray observations - taken in
2013 and 2022 - we see higher maximum velocities of the
blueshifted broad absorption lines in the UV when the X-ray
flux was lower.
} 
    {}

   \maketitle
%
%________________________________________________________________

\section{Introduction}
PG\,0043+039 has been listed as a luminous radio-quiet quasar in the Palomar-Green catalog of ultraviolet-excess stellar objects \citep{green86}  with  $M_V=-26.16$.
An optical spectrum of the H$\beta$ region of PG\,0043+039, taken in September 1990, was presented by \cite{boroson92}. It shows very strong $\ion{Fe}{ii}$ line blends and extremely weak $[\ion{O}{iii}]\,\lambda 5007$ emission. Furthermore, PG\,0043+039
has been identified as a weak broad-absorption-line (BAL) quasar based on the $\ion{C}{iv}$ BAL detected with the Hubble Space Telescope (HST) in 1991 (\citealt{bahcall93}, \citealt{turnshek94, turnshek97},  \citealt{marziani96}).

A ROSAT non-detection established PG\,0043+039 as an X-ray-weak quasar \citep{brandt00}. Non-detections in pointed observations with the ASCA satellite (\citealt{gallagher99}) as well as with XMM-Newton (\citealt{czerny08}) confirmed the extreme X-ray weakness of PG\,0043+039.
 \cite{kollatschny15, kollatschny16} detected PG\,0043+039 in the X-ray based on very deep exposures taken with XMM-Newton in 2013.
PG\,0043+039 exhibited an extreme  $\alpha_{ox}$ gradient
($\alpha_{ox}$=$-$2.37). 
A reanalysis of XMM-Newton X-ray data that were taken in 2005 
gave a similar extreme gradient, with $\alpha_{ox}$=$-$2.53 for that year.
The X-ray spectrum of 2013 was compatible with a normal quasar power-law spectrum and only moderate intrinsic absorption 
(N$_H$~$=$~$5.5_{-3.9}^{+6.9}$~$\times$10$^{21}$~cm$^{-2}$) and reflection. 
The most extreme $\alpha_{ox}$ gradient of PG\,0043+039 and its extreme X-ray faintness -- in comparison to other quasar samples -- were discussed in \cite{kollatschny16}. 

The X-ray faintness of PG\,0043+039
may have been caused by fluctuations in the accretion process,
as seen in low state observations of other AGNs as 
such as PG\,2112+059 \citep{schartel10} or 1H0707-495 \citep{fabian12}. 
Furthermore, the X-ray spectral characteristics can also change because of
changes of the column density and/or the ionization state of the
X-ray-absorbing material. Examples are NGC\,1365 
\citep{risaliti05} and NGC\,7583 \citep{bianchi09}.
Even the optical spectroscopic type of an AGN can
change with time, as observed in changing-look galaxies such as Fairall\,9 \citep{kollatschny85}, NGC\,2617 
\citep{shappee14}, SDSS-AGN \citep{LaMassa15}, and IRAS\,23226-3843 \citep{kollatschny23}. 

The UV/optical flux of PG\,0043+039 increased by a factor of 1.8 in 2013
compared to spectra taken from 1990-1991 \citep{kollatschny16}. The FUV spectrum of 2013 was highly peculiar and dominated by broad bumps and/or absorption troughs besides Ly$\alpha$. 
 There was no detectable Lyman
edge associated with the BAL absorbing gas.
PG\,0043+039 showed a maximum in the UV/optical continuum flux at around 
$\lambda  \approx 2500$ \AA{}  
in contrast to most other AGNs, where no maximum was found in 
this wavelength range.\\

X-ray-weak quasars often show broad absorption lines in the UV 
(\citealt{green95},  \citealt{brandt00}). Broad absorption-line systems may vary with time, as reported for WPVS\,007 \citep{leighly09}, for example. The change in the broad
UV absorption system was reflected in changes of the X-ray brightness in this object.
Broad absorption-line quasars (BALQ) are divided into two categories: high-ionization broad absorption-line quasars (HiBALQs), showing blueshifted absorption from high-ionization lines such as $\ion{C}{iv}\,\lambda 1550$, $\ion{N}{v}\,\lambda 1243$, and  
$\ion{Si}{iv}\,\lambda 1403$ (in order of decreasing typical strength); and
low-ionization broad absorption-line quasars (LoBALQs) showing low-ionization absorption lines such as $\ion{Mg}{ii}\,\lambda 2800$ and/or
 $\ion{Al}{iii}\,\lambda 1863$ in addition to the high-ionization lines (\citealt{hall02},    \citealt{hamann19}, \citealt{choi20}, and references therein).
An extreme subclass of HiBALQs are the $\ion{P}{v}$  BALs exhibiting strong  
$\ion{P}{v}\,\lambda 1128$ absorption (\citealt{capellupo17}, \citealt{hamann19}). Blueshifts and high equivalent widths of highly ionized absorption lines - especially of $\ion{P}{v}$ outflows - are indications of strong galactic winds and high kinetic power in Quasar/ULIRG evolutionary studies \citep{veilleux22}. Many of these objects are X-ray-weak sources.

Here, we reinvestigated the X-ray weakness and/or X-ray variability
of PG\,0043+039 based on new and deep XMM-Newton \& NuSTAR observations taken in 2022.
Furthermore, we studied the highly peculiar UV spectra and their variations based on additional HST observations of this BAL quasar simultaneously to the X-ray observations. These observations were accompanied by new optical spectra.
Finally, we investigated the combined optical-UV spectral energy distributions (SEDs) and their variations.

We took a redshift of z=0.38463$\pm$0.00008.
The coordinates of PG\,0043+039 are RA~=~00:45:47.230 and DEC~=~+04:10:23.40 (2000). 
A luminosity of
$\nu\,L_{\nu} = 2.21 \times10^{44}\,$erg$\,$s$^{-1}$ at 3000 \AA\
has previously been determined for this 
quasar based on data from 1991 \citep{baskin04}. This flux corresponds to an Eddington luminosity $L/L_\text{edd}$ = 0.225 for a black-hole mass of 
$M = 8.9 \times 10^{8} M_{\odot}$ \citep{baskin05}.
Throughout this paper, we assume a $\Lambda$CDM cosmology with a Hubble constant of $H_0$~=~69.6~km s$^{-1}$ Mpc$^{-1}$, $\Omega_{\Lambda}$~=~0.714, and $\Omega_{\rm M}$~=~0.286 from \cite{bennett14}. With a redshift of $z=0.385$, this results in a luminosity distance of $D_{\rm L}$~=~1512.5\,Mpc~=~4.67 $\times 10^{27}$~cm using the cosmology calculator developed by \cite{wright06}. 

The paper is structured as follows. In Sect.~\ref{sec:observations_reduction}, we describe the observations and data reduction. The results of our spectral and continuum variations in the UV,
optical, and X-ray are presented in Sect.~\ref{sec:results}. We discuss these results in Sect.~\ref{sec:discussion} and give a summary in Sect.~\ref{sec:summary}.

\section{Observations and data reduction}\label{sec:observations_reduction}

\subsection{XMM-Newton observations}\label{X-ray_observations}

The XMM-Newton observatory 
\citep{jansen01} conducted two observations of the source 
PG\,0043+039 in the year 2022. The initial observation commenced on June 15, while the subsequent observation began on July 9.

The observation that commenced on June 15 with the observation identifier (ObsId) 9002303 was scheduled to last for a duration of 29.4 ks.
The X-ray instruments, pn-camera \citep{strueder01},
and MOS-cameras \citep{turner01}  were forced to cease operation prematurely due to the presence of elevated radiation levels after approximately two-thirds of the scheduled time had elapsed.
The observation initiated on July 9 with ObsId  9002306 was completed as scheduled for 112.2 ks.
Both X-ray observations were significantly affected by elevated levels of background radiation, which resulted in the full scientific buffer being reached, as documented in the observation protocol. 

During the observation initiated on July 9, the optical monitor was deactivated due to the proximity of Jupiter to the target position, which rendered the instrument inoperable.
Pipeline\footnote{\url{https://xmm-tools.cosmos.esa.int/external/xmm_obs_info/odf/data/docs/XMM-SOC-GEN-ICD-0024.pdf}} (compare with \citealt{webb20}) screening for low-background time intervals identified 11.4 ks of suitable data for the central region of the detectors for the observation initiated on June 15, and 87.515 ks for the observation initiated on July 9. The results are employed in the forthcoming analysis.

\subsection{NuSTAR observation}\label{nustar_observations}

The NuSTAR observatory conducted an observation of PG\,0043+039 between
9 July, 2022, 11:41:08, and 10 July, 2022, 21:41:11, under the unique observation number 80801603002. During the observation, 59,912 seconds of effective exposure were achieved for FPMA and 59,273 seconds for FPMB.

\subsection{UV and optical continuum observations with SWIFT}\label{swift_observations}

We could not use the coaligned 30-cm optical/UV telescope (OM) on board the XMM-Newton satellite as Jupiter was too near to our target in July 2022.
Therefore, we took additional simultaneous observations in the UV and optical continuum bands with SWIFT on July 10, 2022. The exposure time was 2.7 ks.

\subsection{UV (COS) spectroscopy with HST }\label{sec:HST_observations}
We observed PG\,0043+039 over two full HST orbits with the COS spectrograph on July 10, 2022. The COS spectrograph offers two independent observing channels.
The first orbit was devoted to a spectrum taken with the far-ultraviolet (FUV) detector, which is sensitive to wavelengths between 900 and 2150\,\AA. The second orbit was devoted to a spectrum taken with the near-ultraviolet (NUV) detector, which is sensitive to wavelengths from 1650 to 3650\,\AA.  

For the first orbit, we used the COS/FUV spectrograph with the G140L grating and a 2.5 arcsec aperture (circular diameter).
This spectral set covers the wavelength range from 1105\,\AA\
to 2150~\AA\ with a resolving power of 2000 at 1500\,\AA{}.
To fill up the wavelength hole produced by the chip gap and reduce
the fixed pattern noise, we split this observation into four separate segments
of 268 s and four different central wavelengths.
The final exposure time of the integrated FUV spectrum was 1072 s.
The observed FUV spectrum corresponds to 800\,\AA{} till 1552~\AA\
in the rest frame of the galaxy.

For the second orbit, we used the COS/NUV spectrograph with the G230L grating and a 2.5 arcsec aperture.
The full NUV spectrum covers the wavelength range from 1671\,\AA\
to 3642~\AA\ with a mean resolving power of 3000.
Obtaining the full NUV spectrum with G230L requires four CENWAVE settings at 2635, 2950, 3000, and 3360~\AA. Again, we split the observations of each setting into four separate segments of 22 s for reducing the fixed pattern noise.   
Finally, we combined all the individual NUV spectra into one NUV spectrum with an integrated exposure time of 352 s. 
The observed NUV spectrum from 1671\,\AA\ to 3642~\AA\ corresponds to 1207\,\AA{} to 2630~\AA\ in the rest frame of the galaxy.

All the original raw data were processed using the
standard COS calibration pipeline CalCOS. 
We corrected the UV spectra as well as our optical spectra of PG\,0043+039 for Galactic extinction.
We used a reddening value of E(B-V) = 0.018 deduced from
the \cite{schlafly11} re-calibration of the \cite{schlegel98} infrared-based dust map. Furthermore, we applied the reddening law of \cite{fitzpatrick99} with R$_{V}$\,=\,3.1
to our UV/optical spectra.

\subsection{Optical spectroscopy with the HET telescope
}\label{sec:spectroscopic_observations}
We took new optical spectra of PG\,0043+039 with the
9.2m Hobby-Eberly Telescope (HET) at McDonald Observatory
on October 16, 2019; July 9, 2022; and July 17, 2024.
The spectra were taken with the blue unit of the second-generation Low-Resolution Spectrograph \citep[LRS2-B;][]{chonis_2016} attached to the HET.
  
LRS2 is an integral field spectrograph with four spectral channels covering 3640\,\AA\ to 10500\,\AA\ in separate blue and red units (LRS2-B and LRS2-R, respectively), each fed by lenslet-coupled 
fiber internal field units (IFUs). Each IFU covers $12\arcsec \times 6\arcsec$ with 0.6\arcsec spatial elements and full fill-factor, and couples to 280 fibers. The IFUs are separated by 
100\arcsec\ on sky. For the present study, we only utilized the LRS2-B unit, which simultaneously covers the wavelength ranges from 3640 to 4670\,\AA\ and 4540 to 7000\,\AA\ (referred to as the UV and Orange channels, respectively) at fixed resolving powers
of $R \sim 2500$  and  1400, respectively. 
 Table\,\ref{tab:log_of_HET_obs} gives the log of our spectroscopic observations. All spectra  were
taken with identical instrumental setups and  at the same airmass, owing to the particular design of the HET with fixed altitude.

\begin{table}[!h]
\tabcolsep+6mm
\caption{Log of optical spectroscopic observations of PG\,0043+039 taken with HET.}
\centering
\begin{tabular}{ccc}
\hline \hline 
\noalign{\smallskip}
Mod. Julian Date & UT Date & Exp. time \\
            &         &  [sec.]   \\
\noalign{\smallskip}
\hline 
\noalign{\smallskip}
58772.19  &  2019-10-16  & 1400 \\
59769.45  &  2022-07-09  &  787 \\ 
60508.43  &  2024-07-17  &  812  \\
\hline 
\vspace{-.7cm}
\end{tabular}
\label{tab:log_of_HET_obs}
\end{table}

The 280 fiber spectra in each of the UV and Orange channels of LRS2-B were reduced with the automatic HET pipeline, Panacea\footnote{https://github.com/grzeimann/Panacea}. 
This pipeline performs basic CCD reduction tasks, wavelength calibration, fiber extraction, sky subtraction, and flux
calibration.  

All wavelengths were converted to the rest frame of the
galaxy at z=0.38463$\pm$0.00008. This redshift is based on a reanalysis of the narrow
$[\ion{O}{ii}]\,\lambda3727$ line in the spectrum from 2013 when the galaxy was in its highest state. We would obtain a redshift of z=0.38503$\pm$0.00008 if we used the broad $\ion{Mg}{ii}\,\lambda 2800$ line as our reference.
\cite{ho09} derived a redshift of z=0.38512 ± 0.000009 based on a spectrum that they took in 2004.

\section{Data analysis}\label{sec:dataanalysis}

Here, we present the results of our observing campaign based on the data obtained in the
 X-ray, UV, and optical frequency bands.

\subsection{XMM-Newton  }\label{sec:xmm_results}

The XMM-Newton observation conducted on June 15, 2024 yielded no detections of the X-ray source. 
Table~\ref{tab:XMMJunX} presents the two-sigma upper limits for three energy bands and the three detectors.
The sources were identified with a high level of statistical significance in the optical monitor. For reference purposes, the corresponding fluxes are presented in Table~\ref{tab:XMMJunO}.

\begin{table}[!h]
\tabcolsep+3mm
\caption{2-sigma X-ray upper limits for 2022-06-15 XMM-Newton observation.}
\centering
\begin{tabular}{lccc}
\hline \hline 
\noalign{\smallskip}
 band          & \multicolumn{3}{c}{flux} \\
  $keV$      & pn & Mos1 & Mos2     \\
\noalign{\smallskip}
\hline 
\noalign{\smallskip}
0.2 - 12.0    & $<$2.09 & $<$4.75  &  $<$2.71 \\
2.0 - 12.0  & $<$2.89 & $<$5.23  &  $<$3.54\\
0.2 - 2.0     & $<$0.86 & $<$1.78  &  $<$0.97\\
\hline 
\vspace{-.5cm}
\end{tabular}
\tablefoot{Upper limits are in units of 10$^{-14}$\,erg\,s$^{-1}$\,cm$^{-2}$.}
\label{tab:XMMJunX}
\end{table}

\begin{table}[!h]
\tabcolsep+3mm
\caption{XMM-Newton OM fluxes on 2022-06-15.}
\centering
\begin{tabular}{lcc}
\hline \hline 
\noalign{\smallskip}
 filter & central wavelength &  flux \\
 &  [\AA{}]  & [erg\,s$^{-1}$\,cm$^{-2}$\,\AA$^{-1}$]  \\
\noalign{\smallskip}
\hline 
\noalign{\smallskip}
UVW2 & 2120 & 3.009$\pm$0.128 $\times$ 10$^{-15}$\\
UVW1 & 2910 & 2.707$\pm$0.021 $\times$ 10$^{-15}$\\
U    & 3440 & 2.337$\pm$0.018 $\times$ 10$^{-15}$\\
B    & 4500 & 1.396$\pm$0.014 $\times$ 10$^{-15}$\\
V    & 5430 & 0.927$\pm$0.022 $\times$ 10$^{-15}$\\
%\noalign{\smallskip}
\hline 
\end{tabular}
\label{tab:XMMJunO}
\end{table}

In the XMM-Newton observation conducted on 9 July 2024, PG\,0043+039 was unambiguously identified as a weak source.
Table\,\ref{tab:XMMJulX} presents the X-ray fluxes across five energy bands. 
In consideration of the limitations of the source data, 
we proceeded to fit only the pn-spectrum.  
A power law with absorption fixed at N$_H$\;$=$ 2.99 $\times$ 10$^{19}$\;cm$^{-2}$ was assumed \citep{lockman95}. 
Utilizing 38 bins, we obtained a C-value of 31.74.
The resulting values for $\Gamma$ and Norm were 1.46$^{+0.74}_{-0.57}$ and 7.34$^{+2.81}_{-2.65}$\, {photons}\, {keV}\,s$^{-1}$\,cm$^{-2}$  at 1 {keV}, respectively.
\begin{table}[!h]
\tabcolsep+3mm
\caption{X-ray fluxes for 2022-07-9 XMM-Newton observation.}
\centering
\begin{tabular}{lc}
\hline \hline 
\noalign{\smallskip}
 band          & flux \\
  $keV$      &   [erg\,s$^{-1}$\,cm$^{-2}$  ]   \\
\noalign{\smallskip}
\hline 
\noalign{\smallskip}
0.2 - 12.0  &   7.30$\pm$1.56 $\times$ 10$^{-15}$\\
0.5 - 1.0   &   9.19$\pm$1.53 $\times$ 10$^{-16}$  \\
1.0 - 2.0  &    6.40$\pm$1.66 $\times$ 10$^{-16}$  \\
2.0 - 4.5  &    2.42$\pm$0.48 $\times$ 10$^{-15}$  \\    
4.5 - 12.0  &   3.13$\pm$1.46 $\times$ 10$^{-15}$  \\    
\hline 
\vspace{-.7cm}
\end{tabular}
\label{tab:XMMJulX}
\end{table}

\subsection{NuSTAR}\label{sec:nustar_results}

The 2$\sigma$ upper limits for the NuSTAR observation at the optical position of PG\,0043+039 were determined. 
It was assumed that the source region would be circular with a radius of 30 arcsec, while the background region would be circular with a radius of 90 arcsec, situated at a distance of 3.0 arcmin from the source.
Assuming a power law with an index of 2, the 2$\sigma$ upper limit for each detector is 4.5 $\times$ 10$^{-7}$ photons\,s$^{-1}$, which corresponds to 1.0 $\times$ 10$^{-14}$\,erg\,s$^{-1}$\,cm$^{-2}$ 
for the energy range from 10 to 20 keV.

\subsection{SWIFT}\label{sec:swift_results}

The derived fluxes in the SWIFT bands are presented in Tab.\,\ref{tab:SWIFT_obs}.
\begin{table}[!h]
\tabcolsep+3mm
\caption{SWIFT fluxes on 2022-07-10.}
\centering
\begin{tabular}{lcc}
\hline \hline 
\noalign{\smallskip}
 filter & central wavelength &  flux  \\
    &  [\AA{}] &     [erg\,s$^{-1}$\,cm$^{-2}$\,\AA$^{-1}$] \\
\noalign{\smallskip}
\hline 
\noalign{\smallskip}
UVW2 & 1928 & 3.058$\pm$0.054 $\times$ 10$^{-15}$\\
UVM2 & 2246 & 2.837$\pm$0.073 $\times$ 10$^{-15}$\\
UVW1 & 2600 & 2.785$\pm$0.077 $\times$ 10$^{-15}$\\
U    & 3465 & 2.560$\pm$0.046 $\times$ 10$^{-15}$\\
B    & 4392 & 1.771$\pm$0.038 $\times$ 10$^{-15}$\\
V    & 5468 & 1.128$\pm$0.043 $\times$ 10$^{-15}$\\
%\noalign{\smallskip}
\hline 
\end{tabular}
\label{tab:SWIFT_obs}
\end{table}
The central wavelengths of the SWIFT UVOT filters are given in \cite{poole08}.
The SWIFT filter curves are not exactly the same as those of the XMM-Newton OM telescope. However, the measured flux values in the UV and optical bands are identical within a few percent for the observations taken on 2022-06-15 and 2022-07-10. Furthermore, the flux values agree with those based on the UV/optical spectra.

\subsection{UV spectral observations with HST in 2022 }\label{sec:hst_results}

\begin{figure*}
\sidecaption
        \includegraphics[width=12cm, angle=0]{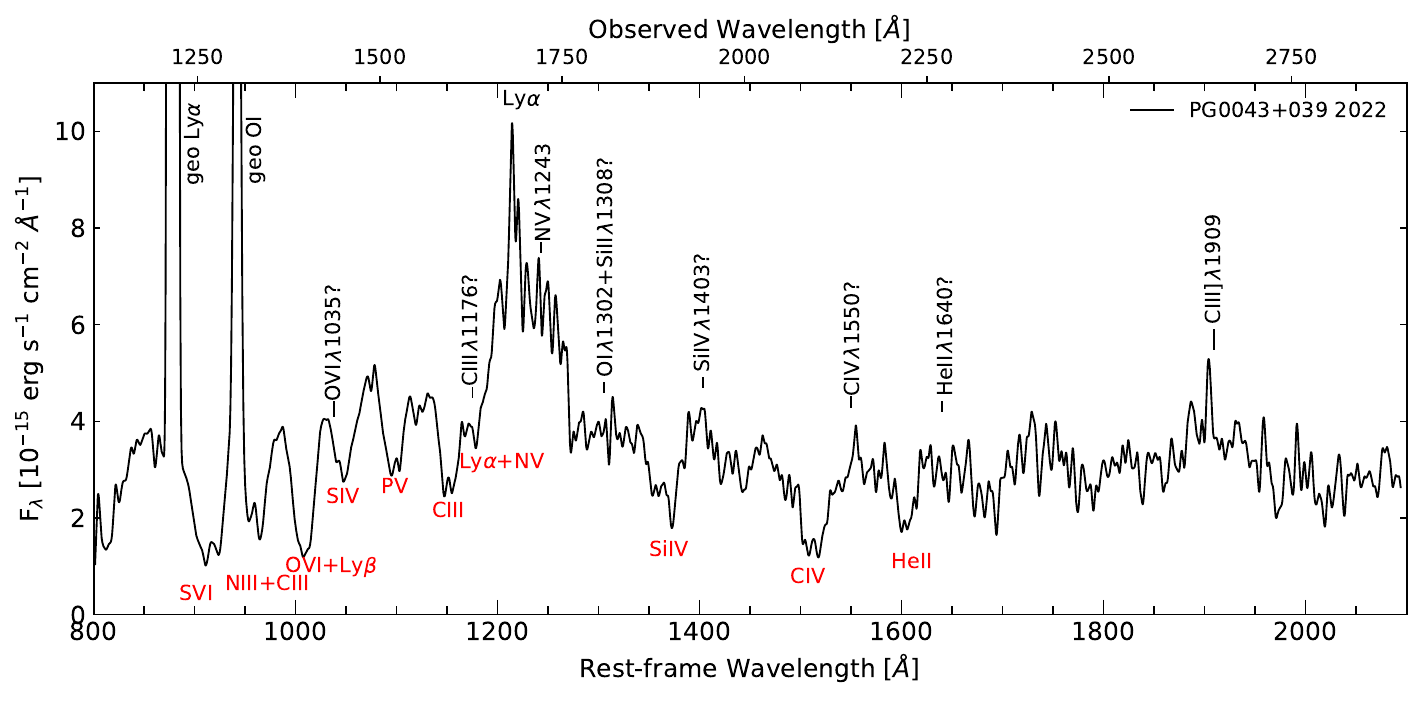}
        \caption{Hubble Space Telescope UV spectrum of PG\,0043+039 taken in July, 2022. The location of possible emission lines is indicated above the spectrum (in black). Absorption troughs are indicated below the spectrum (in red).}
        \label{PG0043_UV_22_20240813}
 \end{figure*}

The combined NUV and FUV spectra of PG\,0043+039 -- obtained in July 2022 with the HST -- are presented in Fig.~\ref{PG0043_UV_22_20240813}. The HST spectrum has been smoothed by means of a running mean width
of $\Delta \lambda = 2.0$ \AA{} for highlighting
weaker spectral structures.  
We do not show the reddest G230L segment (central wavelength at 3360 \AA\ in observed frame) here because of its low S/N ratio.
This wavelength segment (2090 -- 2600 \AA{} in rest frame) is additionally presented in Fig.~\ref{PG0043_UV_opt_22c_pwlaw_20250214} together with the optical spectrum. This segment had to be smoothed by means of a running mean width of $\Delta \lambda = 15.$ \AA{} because of its low S/N ratio.
We indicate the location of possible emission lines,
the geo-coronal lines Ly$\alpha$ and $\ion{O}{vi}$, as well as the position of possible broad absorption troughs in Fig.~\ref{PG0043_UV_22_20240813}. 
Only four emission lines (Ly$\alpha$, \ion{N}{V}\,$\lambda$1243, $\ion{Si}{iv}\,\lambda 1403$, and \ion{C}{iii}]\,$\lambda$1909) can be unambiguously identified, 
besides the two geocoronal lines. It is unclear what the reason is for the emission redward of the \ion{N}{V}\,$\lambda$1243 line.
The emission lines as well as the absorption troughs and their variations are discussed in Section \ref{sec:discussion}.

\subsection{Optical spectra in 2022}\label{sec:optspec_results}

\begin{figure*}
\sidecaption
\includegraphics[width=12cm, angle=0]{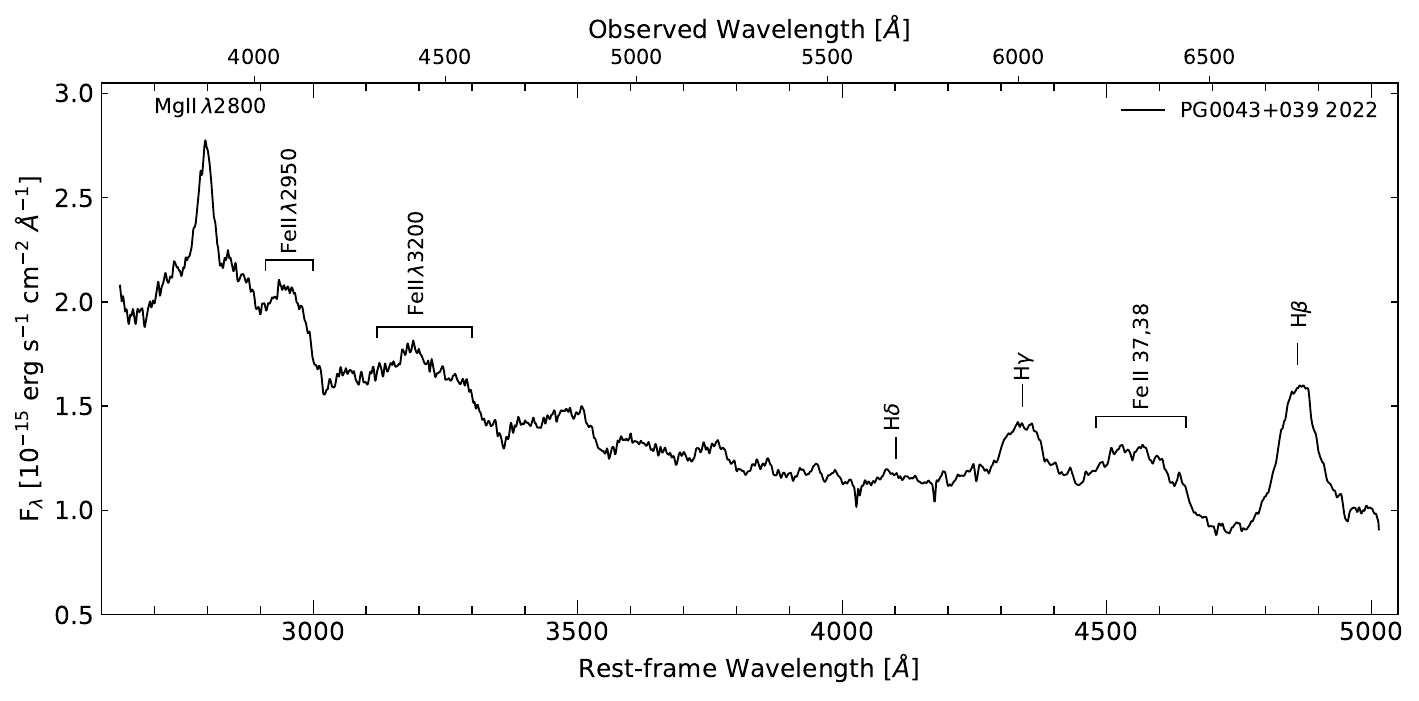}
        \caption{Observed optical spectrum of PG\,0043+039 from July, 2022}
        \label{PG0043_opt_22_20240813}.
 \end{figure*}

The combined optical spectrum -- obtained with the HET telescope in July, 2022  -- is shown in Fig.~\ref{PG0043_opt_22_20240813}. 
The observed spectrum based on the UV and Orange channels of the LRS2-B spectrograph covers the wavelength range of 3750 -- 6800 \AA\ in the observed frame and 2650 --
5020 \AA\ in the rest-frame system. The spectrum is dominated by the broad Balmer lines H$\beta$, H$\gamma$, and H$\delta$; a narrower $\ion{Mg}{ii}\,\lambda$2800 line; and strong $\ion{Fe}{ii}$ blends. One can identify the strong $\ion{Fe}{ii}$ blend between 4450 and 4700 \AA\ based on the $\ion{Fe}{ii}$ multiplets 37, 38, and 43, as well as many bumps 
between 2600 and 4000 \AA\ - for example at 2950 and 3200 \AA\ --
caused by blends of many thousand $\ion{Fe}{ii}$ lines 
(e.g., \citealt{sigut03}).

\subsection{Combined optical UV spectrum in 2022}\label{sec:disc_optuvvar}

\begin{figure*}
\sidecaption
        \includegraphics[width=12cm, angle=0]{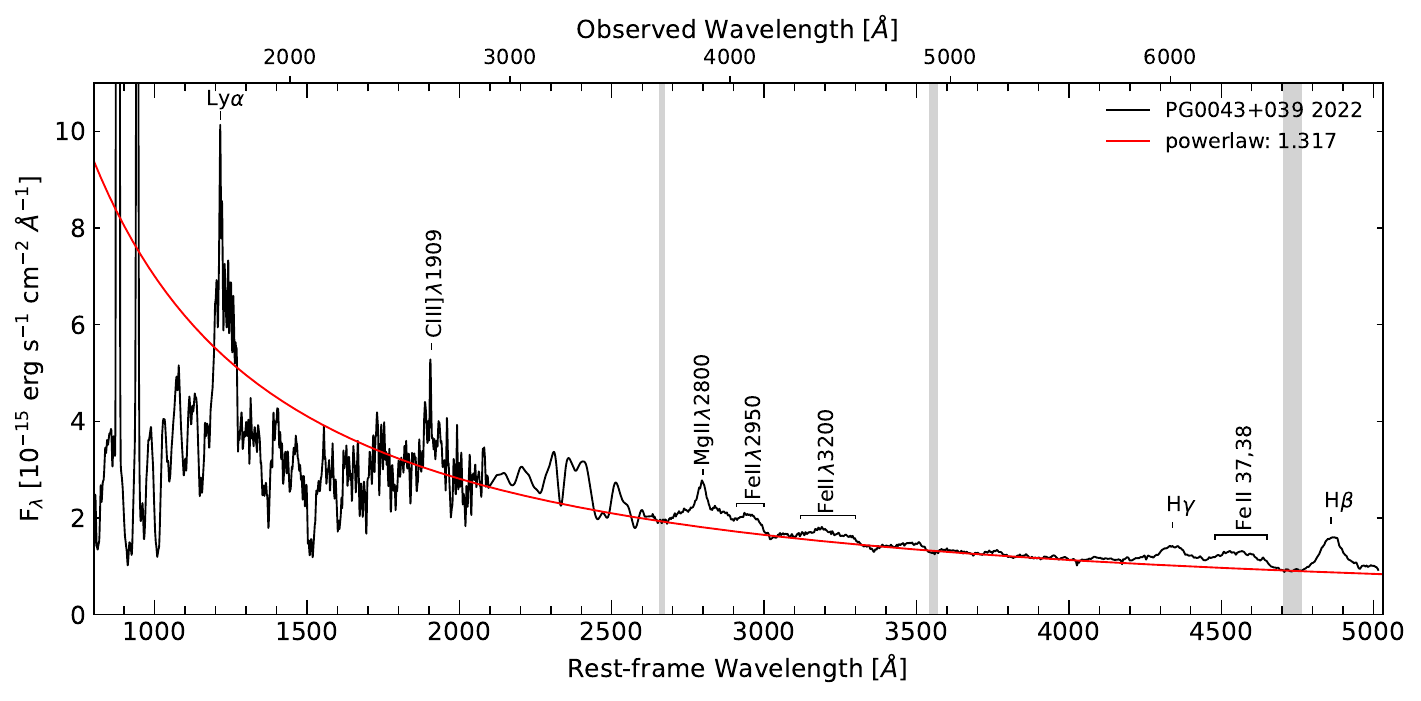}
        \caption{Combined UV-optical spectrum of PG\,0043+039 in 2022. 
        A power-law model (in red) was fit to the continuum, giving a spectral index of $\alpha=1.317$. The spectral regions that were used for deriving the power-law slope of the continuum are shown in gray.}
        \label{PG0043_UV_opt_22c_pwlaw_20250214}
\end{figure*}

 We present a combined optical-UV spectrum of PG\,0043+039 -- based on the data taken in July 2022 -- in Fig.~\ref{PG0043_UV_opt_22c_pwlaw_20250214}. 
We tried to fit a power law to the UV optical continuum. Because of the strong absorption troughs in combination with overlapping broad emission lines in the UV it is demanding to find continuum regions that
are not contaminated. Finally, we used three continuum regions at 
2666$\pm$10, 3555$\pm$15, and 4734$\pm$30 \AA\
to fit a power law to the continuum spectrum
(see Fig.~\ref{PG0043_UV_opt_22c_pwlaw_20250214}). 
We obtain a gradient of $\alpha$ = 1.317 with
 $f_\lambda \sim\lambda^{-\alpha}$. This fit is a first-order approximation to the optical/near-UV continuum gradient of PG\,0043+039 in 2022.

\section{Results}\label{sec:results}

\subsection{Continuum and emission-line variations in the optical}\label{sec:optlinconvar}

\begin{figure*}
\sidecaption
        \includegraphics[width=12cm, angle=0]{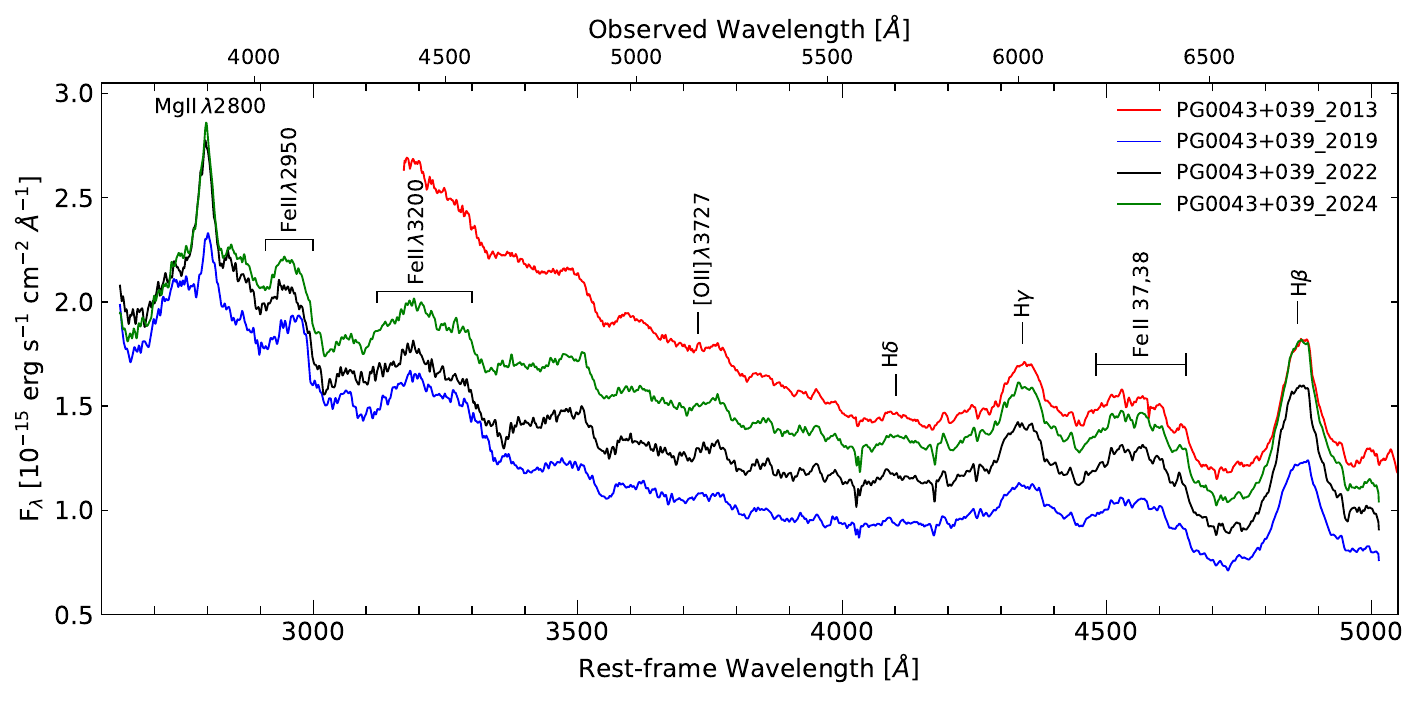}
        \caption{Optical spectra of PG\,0043+039 taken in 2022, as well as in the years 2013, 2019, and 2024.}
        \label{PG0043_optall_13_19_22_24_20240808}
\end{figure*}

In Fig.~\ref{PG0043_optall_13_19_22_24_20240808}, we show optical spectra of PG\,0043+039 taken with the HET telescope in the years 2013, 2022, 2019, and 2024. 
One can see clear variations in the continuum flux, in the continuum gradient, in the Balmer and $\ion{Fe}{ii}$ intensities, as well as in the 
\ion{Mg}{ii}\,$\lambda$2800 line flux and line profile.
\ion{Mg}{ii}\,$\lambda$2800 evidently shows a narrower line profile than the Balmer lines.
The line profiles are discussed in more detail in Section\,\ref{sec:emlinprof}.

Table.\ref{lineintensities} gives the optical continuum intensities at 3552 and 4734\,\AA\ rest frame, as well as the 
Balmer, $\ion{Fe}{ii}$(37,38), and  \ion{Mg}{ii}\,$\lambda$2800 intensities for the year 2022, and furthermore for the years 2013, 
2019, and 2024. In addition, we list the integration limits in column 7.
\begin{table*}
%\centering
\tabcolsep+2.8mm
\caption{Emission line and continuum intensities in rest frame and corrected for Galactic extinction for the years 1991 up to 2024.}
\begin{tabular}{lcccccc}
\hline 
\noalign{\smallskip}
Emission line                 &Flux 91     & Flux 13  &  Flux 19 & Flux 22 & Flux 24      & Wavelength range\\ 
and continuum                              &           &           &          &         &              &    [\AA{}]  \\
\noalign{\smallskip}
(1)                           & (2)     &  (3)      &    (4)              &        (5)  &         (6)                 &  (7)\\
\noalign{\smallskip}
\hline 
\noalign{\smallskip}
cont. 1288                   & 7.60$\pm{}$.3  & 8.03$\pm{}$.3  &   & 6.24$\pm{}$.3    &      & 1287 -- 1290\\
cont. 1818                   & 4.07$\pm{}$.2  &                &   & 3.28$\pm{}$.2    &      & 1808 -- 1828\\
cont. 3552                   &                & 1.89$\pm{}$.2 & 1.06$\pm{}$.2  & 1.28$\pm{}$.2 & 1.55$\pm{}$.2 & 3542 -- 3562\\
cont. 4734                   &                & 1.19$\pm{}$.2 & 0.74$\pm{}$.2  & 0.90$\pm{}$.2 & 1.05$\pm{}$.2 & 4724 -- 4744\\
\noalign{\smallskip}
\hline 
\noalign{\smallskip}
$\ion{O}{vi}\,\lambda 1038$   &         & 39.8$\pm{}$5. &       & $\le$ 5. &  & 1024 -- 1054\\
$\ion{C}{iii}\,\lambda 1176$  &         & 9.35$\pm{}$5. &       & $\le$ 5. &  & 1164 -- 1181\\
Ly$\alpha$                   & 210.$\pm{}$20. & 228.$\pm{}$20.  &     & 167.$\pm{}$15. &  & 1182 -- 1234\\
$\ion{N}{v}\,\lambda 1243$    & 60.$\pm{}$30. & 67.5$\pm{}$20.  &     & 57.7$\pm{}$30. &  & 1234 -- 1253\\
$\ion{Si}{ii}\,\lambda 1308 + \ion{O}{i}\,\lambda 1302$ & 29.8$\pm{}$5. & 66.6$\pm{}$10.  & & $\le$3. &  & 1288 -- 1327\\
$\ion{Si}{iv}\,\lambda 1403$  & 15.0$\pm{}$3. & 16.6$\pm{}$10.  &    & 20.7$\pm{}$6. &  & 1387 -- 1419\\
$\ion{He}{ii}\,\lambda 1640$  & 4.9$\pm{}$2. &          &    & <3. &  & 1631 -- 1651\\
$\ion{C}{iii}]\,\lambda 1909$  & 18.1$\pm{}$8. &       &  & 15.1$\pm{}$10. &  & 1880 -- 1931\\
$\ion{Mg}{ii}\,\lambda 2800$  &    &  & 7.63$\pm{}$.7  & 17.5$\pm{}$1.5 & 16.8$\pm{}$1.5 & 2767 -- 2831\\
$[\ion{O}{ii}]\,\lambda 3727$ &  & 0.33$\pm{}$.04 &   &  &  & 3720 -- 3732\\
H$\gamma$            &  & 33.3$\pm{}$3. & 23.7$\pm{}$3.  & 31.5$\pm{}$3. & 32.5$\pm{}$3. & 4272 -- 4396\\
$\ion{Fe}{ii}\,\lambda\lambda 4500$  &  & 44.8$\pm{}$4. & 41.6$\pm{}$4.  & 52.5$\pm{}$4. & 51.6$\pm{}$4. & 4453 -- 4669\\
H$\beta$             & & 49.2$\pm{}$5. & 41.0$\pm{}$5.  & 63.9$\pm{}$5. & 66.1$\pm{}$5. & 4764 -- 4948\\ 
\noalign{\smallskip}
\hline
\end{tabular}\\
\tablefoot{Continuum flux in units of 10$^{-15}$\,erg\,s$^{-1}$\,cm$^{-2}$\,\AA$^{-1}$.
Line flux in units 10$^{-15}$\,erg\,s$^{-1}$\,cm$^{-2}$.}
\label{lineintensities}
 \end{table*}
For H$\beta$, we subtracted a constant continuum value measured at 4762 \AA\ because we have
no continuum value on the red side of this line as the spectrum ends 
at 5012\,\AA\ in the rest frame, except for the year 2013. One can see in the spectrum taken in 2013 (\citealt{kollatschny16}, Fig. 8) that the continuum value is lower by 10\,\% on the red side compared to the blue side. Furthermore, we now integrate over a smaller wavelength range than in \cite{kollatschny16}. This leads to a general underestimation for the integrated H$\beta$ flux of the order of 30\,\% in 
Tab.\,\ref{lineintensities}. However, the relative H$\beta$ variations for the different epochs are more correct.

For H$\gamma$ and $\ion{Fe}{ii}\,\lambda\lambda 
4500$, we first subtracted a continuum based on the continuum intensity values at 4207 and 4734 \AA\ before integrating 
the line intensities. This procedure gives a good estimation for the relative line variations. The measured H$\gamma$ line intensities are
larger than the intrinsic values because the underlying continuum is contaminated by strong $\ion{Fe}{ii}$ blends. 

\begin{figure*}
\sidecaption
        \includegraphics[width=12cm, angle=0]{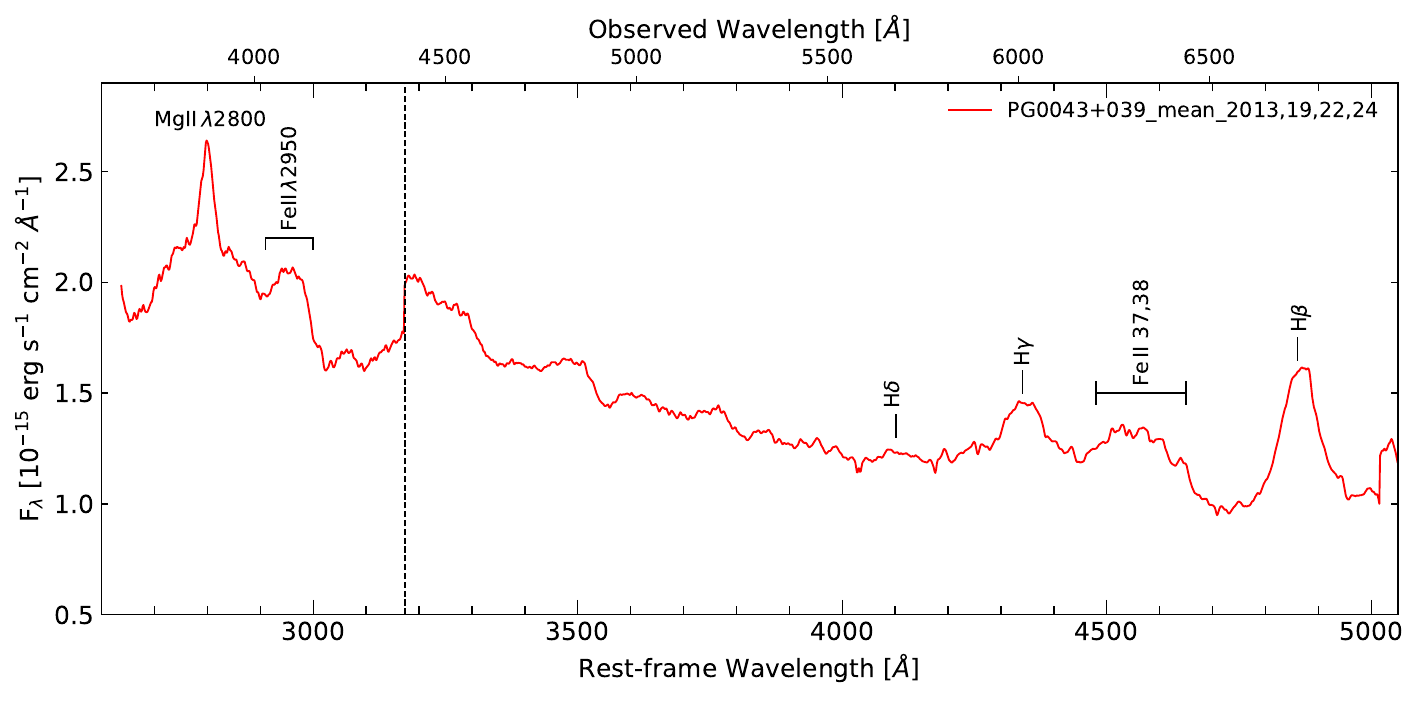}
        \caption{Mean optical spectrum of PG\,0043+039 based on years 2013, 2019, 2022, and 2024. The mean spectrum shortward of 3173 \AA{} (dashed line) is lower because it is based on three spectra only (2019, 2022, 2024).}
        \label{PG0043_opt4mean_13_19_22_24_20241030}
\end{figure*}

Figure~\ref{PG0043_opt4mean_13_19_22_24_20241030} shows the mean optical spectrum of PG\,0043+039 based on the spectra for the years 2013, 2019, 2022, and 2024. The mean spectrum shortward of 
3173\,\AA{} is lower because it is based on only three spectra (2019, 2022, 2024) when PG\,0043+039 was in a lower state.
The optical and UV spectra in PG\,0043+039 down to 
$\sim$ 2400 \AA{} are dominated by strong UV continuum flux and strong $\ion{Fe}{ii}$ line blends.
 See also Figs.\,7 and 13 in
\cite{kollatschny16} and Fig.\,1 in \cite{turnshek94}.
Early observations of the UV bump in quasars as well as its modeling are discussed in, for example, \cite{grandi82} and \cite{wills85}. Recent modeling of the $\ion{Fe}{ii}$ UV emission is presented in \cite{pandey24}.

\begin{figure*}
\sidecaption
        \includegraphics[width=12cm, angle=0]{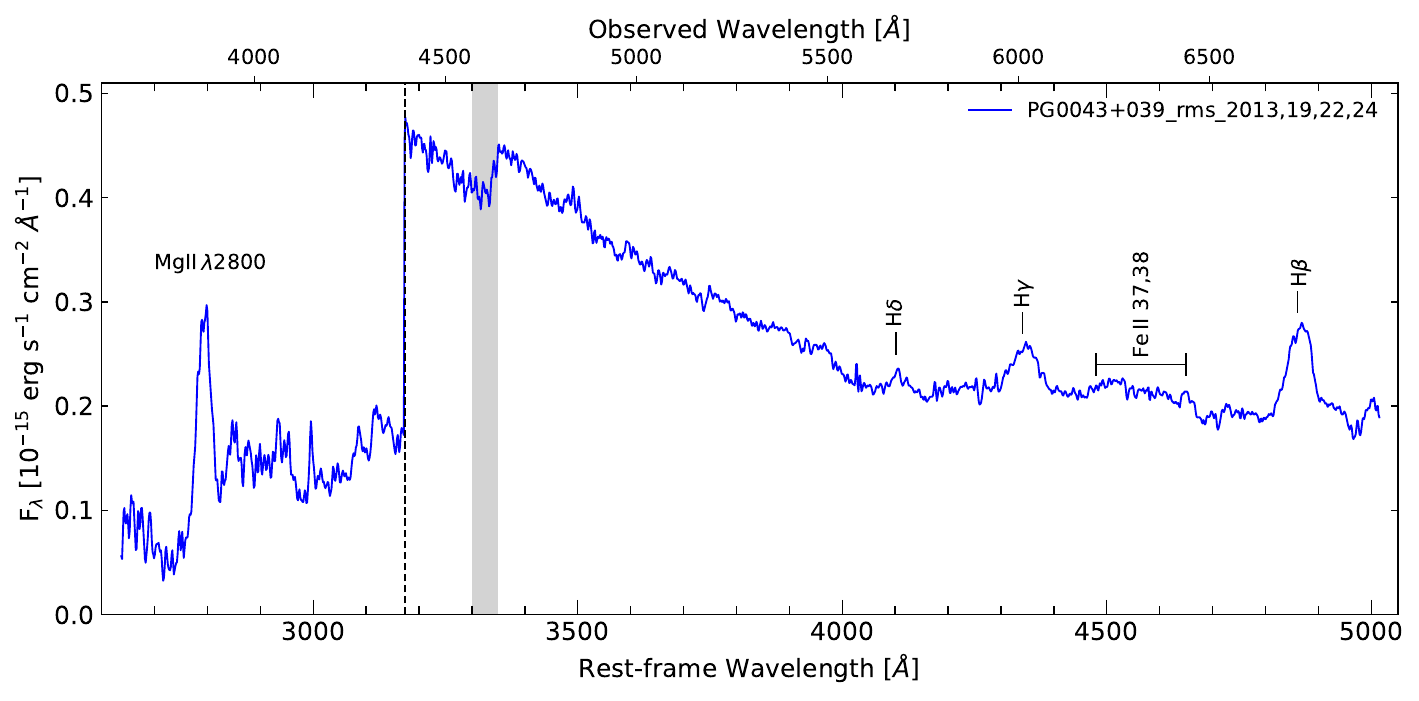}
        \caption{Optical rms spectrum of PG\,0043+039 based on years 2013, 2019, 2022, and 2024. The overlapping region 
        of the UV and Orange channels of the 
        LRS2-B spectrograph is shown in gray. The rms spectrum shortward of 3173 \AA{} (dashed line) is based on three spectra only (2019, 2022, 2024).}
        \label{PG0043_opt_rms_13_19_22_24_20241030}
\end{figure*}

Figure~\ref{PG0043_opt_rms_13_19_22_24_20241030} shows the optical rms spectrum based on the years 2013, 2019, 2022, and 2024. The rms spectrum presents the variable part
of the spectra. One can recognize strong variations in the UV continuum, in the Balmer lines, and especially in the $\ion{Mg}{ii}\,\lambda 2800$ line. 
The $\ion{Fe}{ii}$ blends show only minor variations.
The overlapping region of the UV and Orange channels of the LRS2-B spectrograph is shown in gray. The intercalibration of these two channels was not perfect in our case; i.e., not better than three percent. The relative flux of the UV channel has been slightly underestimated in its original data. 
The sharp drop at 3170\,\AA{} is caused by the fact that the rms spectrum blueward of this wavelength is based on only three spectra of the years 2019, 2022, and 2024,
while the redward spectrum is based on four spectra of the years
2013, 2019, 2022, and 2024.
The strong increase of the variable flux from 4000\,\AA{} to 3170\,\AA{} in Fig.~\ref{PG0043_opt_rms_13_19_22_24_20241030} must be caused by a variable continuum component and not by variations of $\ion{Fe}{ii}$ blends because it rises so smoothly.

\subsection{Continuum and emission-line variations in the UV}\label{sec:uvcontlinvar}

\begin{figure*}
\sidecaption
        \includegraphics[width=12 cm,  angle=0]{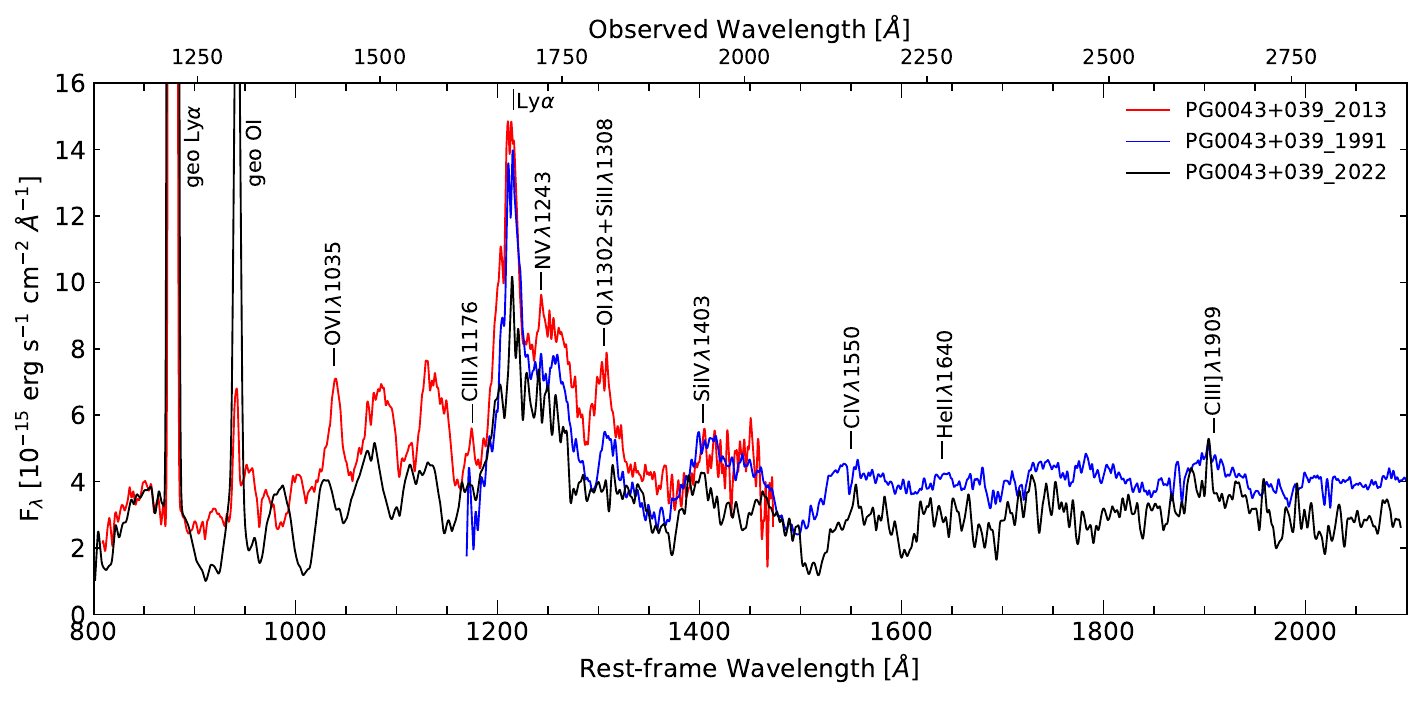}
        \caption{HST UV spectra of PG\,0043+039 taken in years 1991, 2013, and 2022.}
        \label{PG0043_UV_all_20240816}
\end{figure*}

Figure~\ref{PG0043_UV_all_20240816} shows the HST UV spectra up to 2200 \AA\ in the rest frame taken during the years 1991, 2013, and 2022.
The locations of possible emission lines as well as the geocoronal lines are 
indicated above the spectrum. 
The UV spectra have signal-to-noise ratios of 13.9 in 1991, 9.1 in 2013, and 8.9 in 2022  in their continuums at 1420$\pm$20\AA{}.
Besides continuum variations and variable broad emission lines, as seen in the
optical, we also note very broad and variable absorption troughs in the UV. They vary
independently in strength and shape. That makes it very demanding to estimate
the continuum below the broad UV emission lines.
We list the UV continuum intensities at 1288 and 1818 \AA\ rest frame, as well as the UV-emission-line intensities of, for example, Ly$\alpha$ for the years 1991, 2013, and 2022 in Tab.\,\ref{lineintensities}.
In the UV, we rearranged the continuum in comparison to Paper 1. We now take into account stronger absorption troughs. The same is true for the Ly$\alpha$ and
$\ion{N}{v}\,\lambda 1243$ emission lines where we now set the continuum value at 1288 \AA{}.
The line complex $\ion{Si}{ii}\,\lambda 1308 + \ion{O}{i}\,\lambda 1302$ is highly variable with respect to its intensity and wavelength ranges when comparing the spectra from 1991 with those from 2013. This line is not recognizable in the spectrum taken in 2022.
The high ionization $\ion{O}{vi}\,\lambda 1038$ line was strong and clearly present in 2013 when the UV/X-ray flux was strongest. However, it is only marginally detectable in 2022.
The $\ion{O}{vi}\,\lambda 1038$ and $\ion{Si}{iv}\,\lambda 1403$ lines 
decreased by factors of eight or more between 2013 and 2022. The other UV emission lines such as Ly$\alpha$ only decreased by a factor of 1.4. These measurements are affected by major errors because of parallel variations of the absorption troughs. In contrast to that, the optical emission lines exhibited only minor intensity variations between 2013 and 2022.

\begin{figure*}
\sidecaption
        \includegraphics[width=12 cm, angle=0]{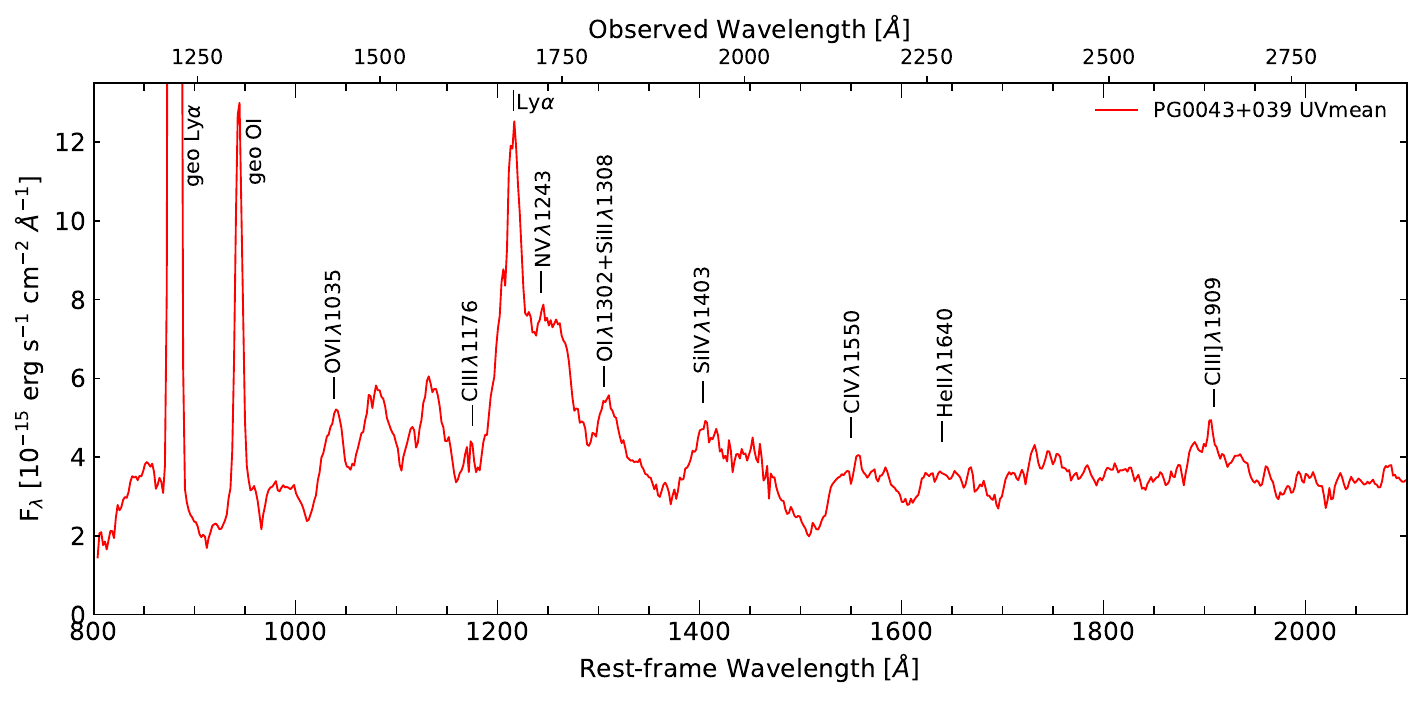}
        \caption{HST UV mean spectrum of PG\,0043+039 for years 1991, 2013, and 2022.}
        \label{PG0043_UV_mean_20240816}
 \end{figure*}

\begin{figure*}
\sidecaption
        \includegraphics[width=12 cm,  angle=0]{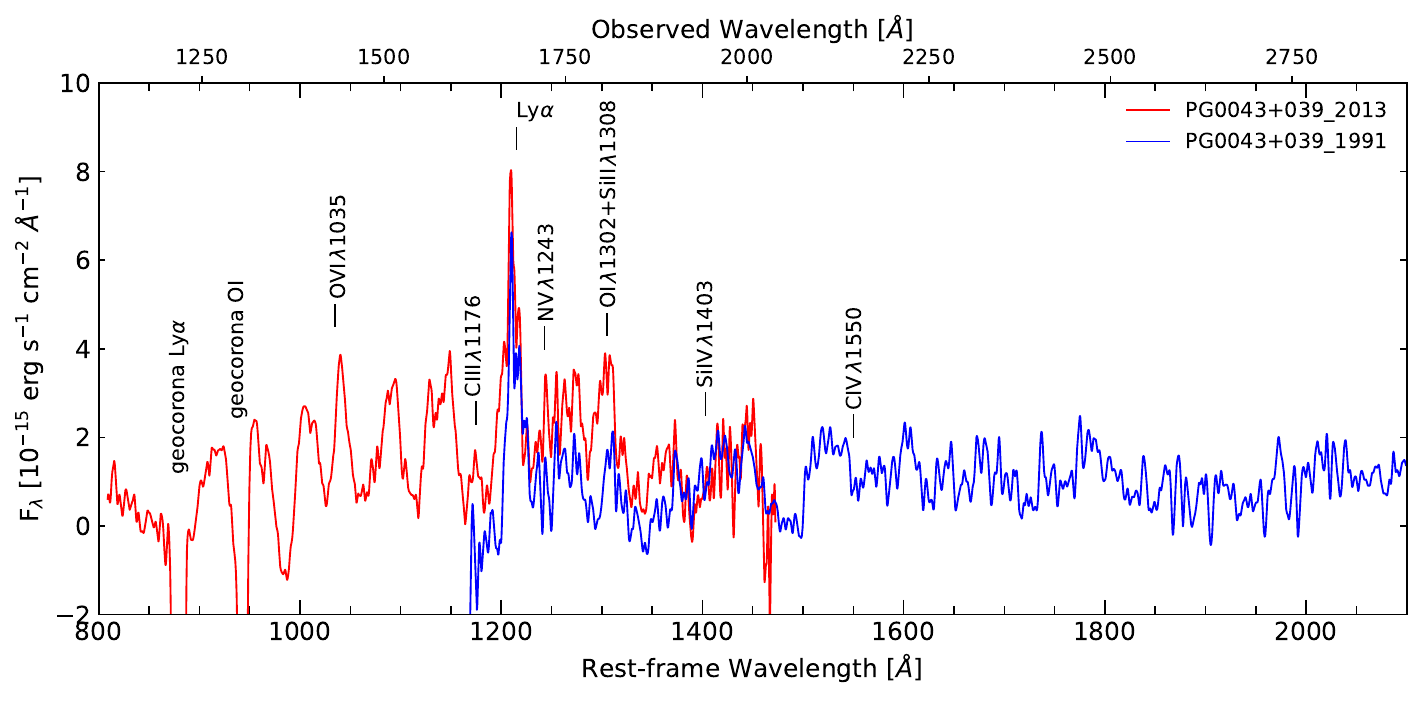}
        \caption{HST UV difference spectra for 1991 and 2013 with respect to the minimum spectrum from 2022.}
        \label{PG0043_UV_diffs_20250620}
\end{figure*}

Figure~\ref{PG0043_UV_mean_20240816} shows the UV mean spectrum of PG\,0043+039 based the HST spectra for the years 1991, 2013, and 2022. The location of possible emission lines is indicated by tick marks above the spectrum.
Figure~\ref{PG0043_UV_diffs_20250620} shows the UV difference spectra for the years 1991 and 2013 with respect to the minimum spectrum from 2022.
The difference spectra demonstrate the 
strong variations of some emission lines in terms of intensity as well as the additional absorption
variability in intensity and velocity. We will discuss this in more detail in Section \ref{sec:optUVcontvar}.

\subsection{Combined optical and UV continuum variations}\label{sec:optUVcontvar}

\begin{figure*}
\sidecaption
        \includegraphics[width=12 cm,  angle=0]{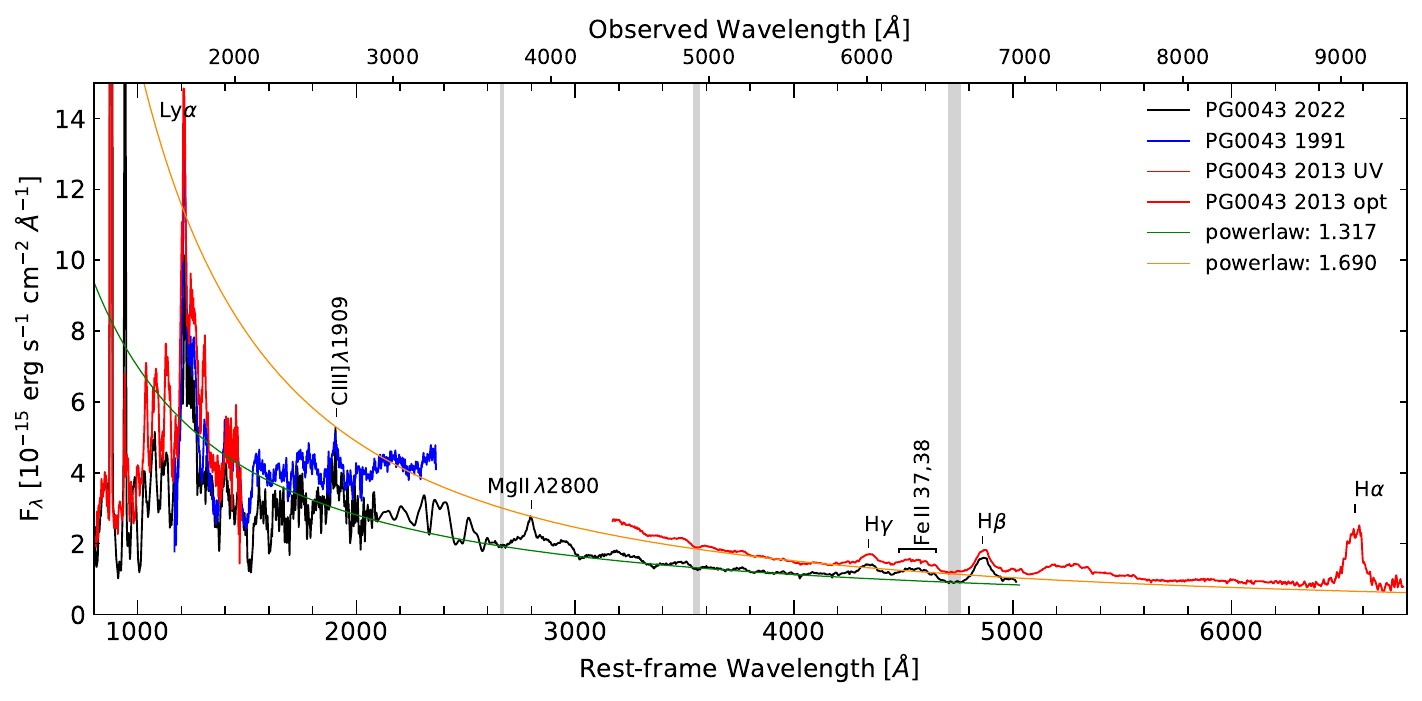}
        \caption{UV and optical spectra of PG\,0043+039 for 2013 and 2022, as well as modeled power-law spectra based on the optical continuum data for 2013 and 2022. In addition, the UV spectrum taken in 1991 is shown.}
        \label{PG0043_UV_opt22d_13_pwlaw_20250224}
\end{figure*}

We present the UV and 
optical spectra of PG\,0043+039 taken in 2013 and 2022 as well as modeled 
power-law spectra based on the optical continuum data for 
these years in Fig.~\ref{PG0043_UV_opt22d_13_pwlaw_20250224}. 
In addition, the UV spectrum taken in 1991 is overlaid. 
Figure~\ref{PG0043_UV_opt22d_13_pwlaw_20250224}
is an extension of Fig.~\ref{PG0043_UV_opt_22c_pwlaw_20250214}. 
The spectral regions that were used for deriving the power-law slope of the continuum are shown in gray.
In principle, we are able to model the combined optical-near-UV continuum 
for the year 2022 with a simple power law $f_\lambda \sim\lambda^{-\alpha}$, where the gradient $\alpha$ has a value of 1.317
(see Section \ref{sec:disc_optuvvar}). PG\,0043+039 was in its lowest state during that year. The slope in the UV is very uncertain due to the strong absorption troughs and emission lines. When we try to fit the optical continuum for the year 2013, when PG\,0043 was in a high state, we obtain 
a gradient of $\alpha$ = 1.69.
However, in the far-UV we observed a negative gradient  for that year (see \citealt{kollatschny16}). The far-UV continuum also exhibited a negative gradient in 1991.

These observations imply not only that the general continuum gradient is changing in the optical, but also that the observed
continuum in the UV changes to an inverted gradient at around 2450 $\pm$ 100 \AA{} (see 
\citealt{kollatschny16}, \citealt{turnshek94}). The wavelength of
this turning point did not change at all from 1991 
\citep{turnshek94} to 2005 based on XMM-Newton OM observations \citep[see Fig.13 in][]{kollatschny16}, or to 2013 \citep{kollatschny16}, and up until 2022, independent
of the different continuum flux levels and continuum gradients.
An evolution in the spectral energy distribution can be seen: when the
optical continuum flux and the gradient were stronger in the years 1991 and 2013, the break and the inverse gradient in the far-UV were
stronger as well. However, when the optical continuum flux and the gradient were flatter in 2022,  the inverse gradient in the 
far-UV was flatter as well. 
This is similar to the general finding of \cite{meusinger16} and others that 3000\,\AA{}
break quasars tend to be intrinsically more luminous than ordinary quasars.

The strong deficit in flux shortward of 2400\,\AA{} might be explained 
with an anomalous steep reddening curve (e.g.,\citealt{turnshek94}, 
\citealt{hall02}, \citealt{veilleux13}, \citealt{kollatschny16}, 
\citealt{meusinger16}, \citealt{choi20}). Another possibility is the existence of an accretion disk with a strongly suppressed
innermost region (\citealt{hall02}, \citealt{kollatschny16}) or a combination of both effects.

In \cite{kollatschny16}, we present the result that the equivalent-width ratios of the strongest UV emission lines ($\ion{O}{vi}\,\lambda1038$, Ly$\alpha$, $\ion{N}{v}\,\lambda1243$) in PG\,0043+039 are similar to those of a composite spectrum of a normal AGN \citep{shull12}.
The Ly$\alpha$ emission line intensity in PG\,0043+039 is always stronger than 
the $\ion{N}{v}\,\lambda1243$ flux for all observations. This behavior of PG\,0043+039 is different with respect to that of all the $\ion{P}{v}$ BALs in the samples 
of \cite{capellupo17} and \cite{hamann19}, where the Ly$\alpha$-emission-line intensity is always significantly weaker than the $\ion{N}{v}\,\lambda1243$ line (see Section \ref{sec:disc_absvar}).

\section{Discussion}\label{sec:discussion}

\subsection{X-ray, UV, optical continuum variations, and $\alpha_{OX}$ variations for the years 2005, 2013, and 2022 \label{sec:contvar}}

We present the X-ray, UV, and optical continuum
intensities in Tab.\,\ref{tab:contvar} to demonstrate their variations from 2013 to 2022 and 
their $\alpha_{ox}$ and $L_{2keV}$ variations.
We also list the X-ray flux and $\alpha_{ox}$ values for 2005 \citep{kollatschny16}.
\begin{table}[!h]
\tabcolsep+1mm
\caption{X-ray, UV, and optical continuum flux variations, as well as $\alpha_{ox}$ and $L_{2keV}$ variations.}
\centering
\begin{tabular}{lcccc}
\hline \hline 
\noalign{\smallskip}
 cont. band, &  2005-06     &   2013-07   &   2022-06    &  (2013/2022) \\
 $\alpha_{ox}$, $L_{2keV}$   &   flux  &   flux   &  flux &  flux ratio\\
\noalign{\smallskip}
\hline 
\noalign{\smallskip}
0.2-12.\,$keV$ & 6.5 $\pm$2.2   & 24.8$\pm$0.28 &  7.30$\pm$1.56 & 3.40 \\
1288\,\AA{} &         & 8.03$\pm$0.3  & 6.24$\pm$0.3 & 1.29 \\
3552\,\AA{} &         & 1.89$\pm$0.2  & 1.28$\pm$0.2 & 1.48 \\
4734\,\AA{} &         & 1.19$\pm$0.2  & 0.90$\pm$0.2 & 1.32 \\
\hline
\noalign{\smallskip}
$\alpha_{ox}$& -2.53$\pm{}$.04  & -2.37$\pm{}$.04  & -2.47$\pm{}$.04 &\\ 
\hline
\noalign{\smallskip}
$L_{2keV}$  & 1.48    &  4.83  &  1.41 & 3.42   \\
\hline 
\end{tabular}
\tablefoot{X-ray flux in units of 10$^{-15}$\,erg\,s$^{-1}$\,cm$^{-2}$.
Optical flux in units of 10$^{-15}$\,erg\,s$^{-1}$\,cm$^{-2}$\,\AA$^{-1}$.
Monochromatic luminosity at 2 keV in units of 10$^{24}$\,erg\,s$^{-1}$\,Hz$^{-1}$.}
\label{tab:contvar}
\end{table}
All flux values decreased from 2013 to 2022. However, the drop was far stronger in the X-ray (a factor of 3.4) than in the optical/UV, where the flux decreased by 25-30\,\%. The relatively low flux in the X-ray with respect to the UV flux -- when the galaxy became fainter -- is consistent with the general $\alpha_{ox}$ - $L_{UV}$ anti-correlation (e.g.,
\cite{wilkes94}, \cite{gibson08b}).

\begin{figure}
        \includegraphics[width=8.5cm, angle=0] {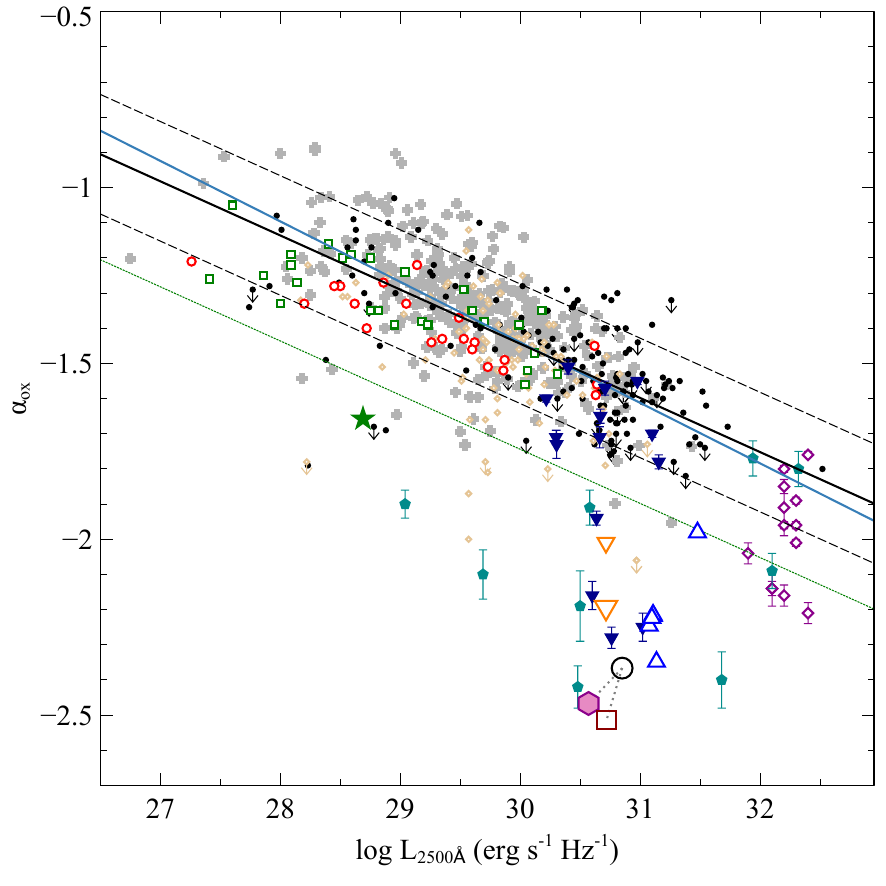}       
        \caption{$\alpha_{ox}$ gradient as function of monochromatic
luminosity at rest frame $2500\,\AA${}. 
Black dots mark the SDSS objects from \cite{strateva05}, while gray filled circles mark the data from \cite{steffen06}.
Further measurements of extreme X-ray weak quasars
are indicated by cyan pentagons (\citealt{saez12}), 
blue and yellow open triangles (\citealt{schartel10}, \citealt{schartel07}, \citealt{ballo11}),
dark blue filled triangles (\citealt{laurenti22}), and magenta open diamonds (\citealt{zappacosta20}).
 The positions of PG\,0043+039 are highlighted for
the years 2005 (red square),
2013 (black circle), and 2022 (filled purple hexagon). 
The solid black line gives the best-fit relation from \cite{lusso10}
with the 1$\sigma$ spread (dashed lines) (see Fig. 5 in 
\citealt{laurenti22}). The blue line is the best-fit relation from
\cite{martocchia17}.  The solid green line marks the line for X-ray 
weakness (\citealt{Pu20}).}
\label{PG0043_Ballo_aox_L2500_2025_6}
\end{figure}

We present  the $\alpha_{ox}$ gradient for samples of AGNs and for PG\,0043+039 as a function of the monochromatic
luminosity at the rest frame 2500\,\AA{} in Fig.~\ref{PG0043_Ballo_aox_L2500_2025_6}. 
Black dots mark the SDSS objects from \cite{strateva05}, while gray filled circles mark the data from \cite{steffen06}.
 Further measurements of extreme X-ray weak quasars
are indicated by cyan pentagons (\citealt{saez12}), 
blue and yellow open triangles (\citealt{schartel10}, \citealt{schartel07}, \citealt{ballo11}), and a green star \citep{ballo08}.
A similar figure -- without the new values from 2022 -- was presented in \cite{kollatschny16}.
Additional data of high-$\lambda_{Edd}$ AGNs
and X-ray-weak quasars are from \cite{laurenti22} (dark blue filled triangles) and \cite{zappacosta20} (magenta open diamonds).
The solid black line gives the best-fit relation from \cite{lusso10}
with the 1$\sigma$ spread (dashed lines) (see Fig. 5 in \citealt{laurenti22}). The blue line is the best-fit relation from
\cite{martocchia17}. The solid green line marks X-ray weakness (\citealt{Pu20}). 

The positions of PG\,0043+039 are highlighted for
the years 2005 (red square),
2013 (black circle), and 2022 (filled purple hexagon). 
PG\,0043+039 shows, among all galaxies, the most extreme $\alpha_{ox}$ gradient
($\alpha_{ox}$=$-$2.53$\pm{}$0.04) in the year 2005.
The value of 
$\alpha_{ox}$=$-$2.37$\pm{}$0.04 increased slightly in the 2013 data. Afterwards, the gradient again decreased to an extreme value of 
$\alpha_{ox}$=$-$2.47$\pm{}$0.04 in 2022. The error of $\alpha_{ox}$ is based on the errors of the X-ray fluxes (Table\,\ref{tab:XMMJulX}) and the errors of the UV fluxes (Table\,\ref{tab:XMMJunO}).
We can certify that the X-ray faintness of PG\,0043+039 is intrinsic and not simulated by variations because of our simultaneous observations in the different frequency ranges.  

\subsection{The Eddington ratio}\label{sec:eddington} 

{\cite{baskin04,baskin05}} derived an Eddington ratio of
$L/L_\text{edd}$ = 0.225 for PG\,0043+039 based on a black-hole mass of  
$M =8.9\times 10^{8} M_{\odot}$ \citep{baskin05}. We now calculate Eddington ratios based on 
the H$\beta$ and \ion{Mg}{ii}\,$\lambda2800$ lines for the year 2022.
We derived black-hole masses of $M=6.95\times 10^{8} M_{\odot}$ (H$\beta$) and 
$M=3.53\times 10^{8} M_{\odot}$ (\ion{Mg}{ii}\,$\lambda2800$) based on the formulas given in \cite{trakhtenbrot12}, on the line widths (FWHM), and the luminosities of $L_{5100} = 1.26 \times 10^{45}\,$erg$\,$s$^{-1}$ and $L_{3000} = 1.42 \times 10^{45}\,$erg$\,$s$^{-1}$.
With a bolometric correction (BC) according to \cite{netzer19}, we derived Eddington ratios of
$L/L_\text{edd}$ = 0.115 (H$\beta$) and 0.157 (\ion{Mg}{ii}\,$\lambda2800$). The derived
Eddington ratios are moderate to high but not as extreme as the 'super-Eddington' objects presented in \cite{laurenti22}. However, we have to keep in mind that our observed luminosities
might be lower limits because the central UV continuum source is partially covered by outflowing gas (see Section \ref{sec:disc_absvar}).

\begin{figure}
        \includegraphics[width=8.5cm, angle=0] {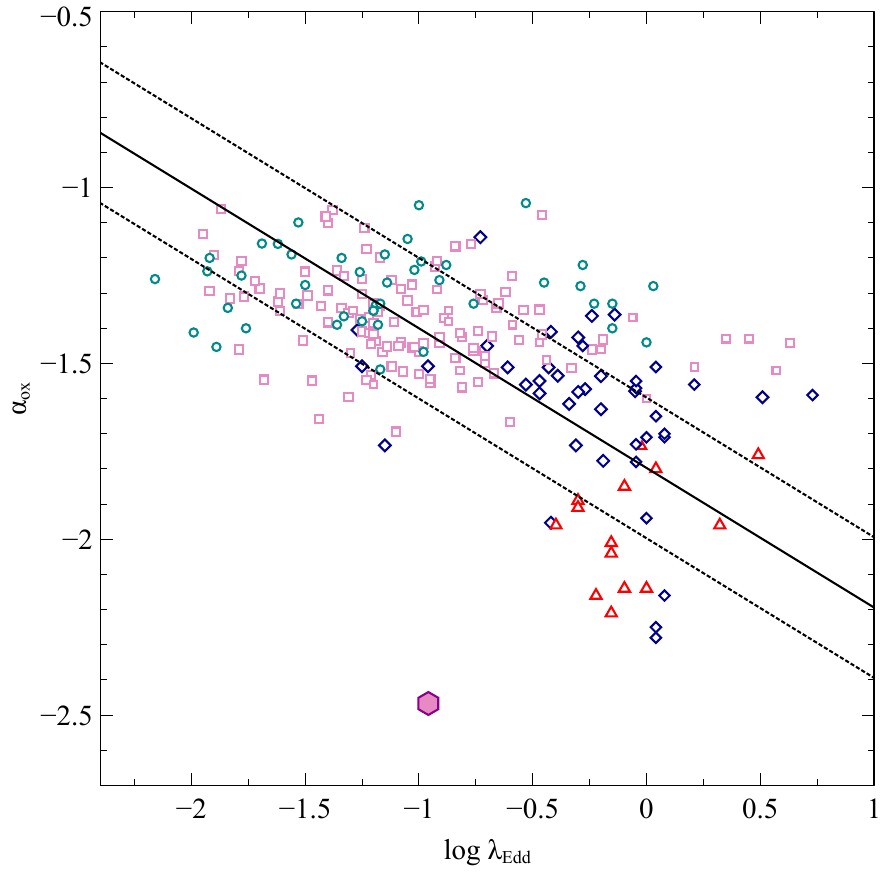}       
        \caption{$\alpha_{ox}$ gradient as function of Eddington luminosity log $\lambda_{Edd}$. 
  The position of PG\,0043+039 (in the year 2022) is highlighted (filled purple hexagon).
The other AGNs are from \cite{laurenti22}, \cite{lusso10}, \cite{liu21}, and \cite{zappacosta20}  divided into different regimes of bolometric luminosity as presented by \cite{laurenti22} (their Fig.\,5, right panel).
The solid line represents the best-fit relation from \cite{lusso10}. The dashed lines give the 1$\sigma$ spread of their relation. }
 \label{PG0043_Ballo_aox_Ledd_2025_10}
\end{figure}

\cite{laurenti22} presented in their work a X-ray spectral analysis and UV/opt photometry of 14 highly accreting AGNs with high Eddington ratios
$L_\text{edd} \gtrapprox$ 1. Their sample is a subsample of high-luminous
galaxies \citep{marziani14} showing extremely strong \ion{Fe}{ii} emission, weak [\ion{O}{iii}] emission, and narrow emission lines with FWHM(H$\beta$) $\le$\,4000\,\kms. The X-ray properties of their sample were quite heterogeneous despite 
the homogeneity of the sample with respect to the high-Eddington ratios. Thirty percent of their sources were 
extremely X-ray weak (see their $\alpha_{ox}$ values in 
Fig.~\ref{PG0043_Ballo_aox_L2500_2025_6}). 
Figure~\ref{PG0043_Ballo_aox_Ledd_2025_10} shows the $\alpha_{ox}$ gradient as a function of the Eddington luminosity log $\lambda_{Edd}$ for a quasar sample
of sub- to super-Eddington AGNs.
The AGN sample is from the work of \cite{laurenti22} (their Fig.\,5, right panel). PG\,0043+039 is a clear outlier in this diagram.

PG\,0043+039 is as well an extremely X-ray weak source and shows very weak [\ion{O}{iii}] emission as well as strong optical
\ion{Fe}{ii} emission with $R_{\ion{Fe}{ii}}$ = EW(\ion{Fe}{ii}$\lambda$\,4570)/EW(H$\beta$) = 0.8. However, it is a 
Pop B object (see \citealt{marziani14}) with respect to its FWHM(H$\beta$) of 5000 \kms.
Furthermore, its Eddington ratio of 0.1 is only moderate with respect to the highly accreting Pop A objects in the sample of \cite{laurenti22} having $R_{\ion{Fe}{ii}} \geq$ 1.

\subsection{UV and optical emission-line profiles}\label{sec:emlinprof}

\begin{figure}
        \includegraphics[width=9.1cm, angle=0] 
        {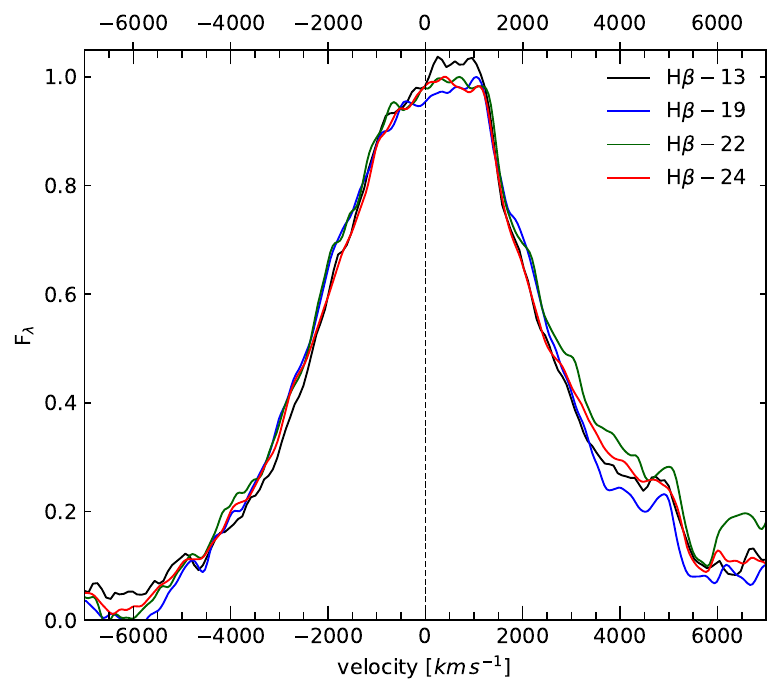}       
        \caption{ Scaled H$\beta$ line profiles for the epochs 2013, 2019, 2022, and 2024}
        \label{Hbprofiles_4epochs}
\end{figure}

We present the scaled H$\beta$ emission line profiles in velocity space for the epochs 2013, 2019, 2022, and 2024 in Fig.~\ref{Hbprofiles_4epochs}.
The H$\beta$ profiles remained constant within a few percent over this period
of time.
The line widths (FWHM) and the shifts of the line centers of all optical and UV emission lines
are listed in Tab.\,\ref{em_linewidths}.
The line shifts are given with respect to the narrow [\ion{O}{ii}]\,$\lambda$3727 line. The narrow [\ion{O}{ii}]\,$\lambda$3727 line could only be verified in the
spectrum taken in 2013 when the galaxy was in its highest state. The
[\ion{O}{iii}]\,$\lambda\lambda$4959,5007 lines are not detectable. Therefore, they
could not be used for the intercalibration of the optical spectra.

\begin{figure*}[t] % Top placement
    \centering
    \hspace*{\fill}%
    % Left side with two stacked figure
    \begin{minipage}[t]{0.46\textwidth}
        \centering
        \vspace{0pt}
        \includegraphics[width=1.0\textwidth]{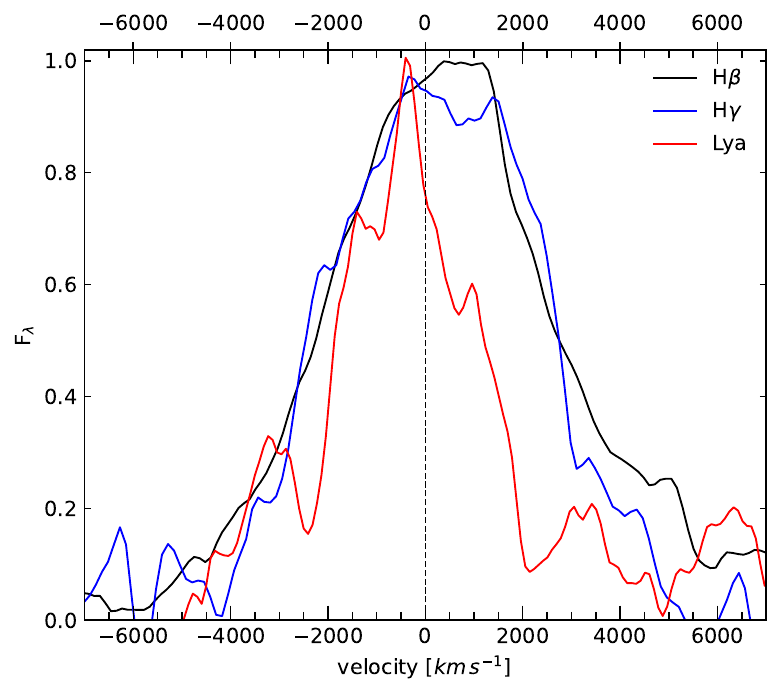}
        \caption{Scaled mean Balmer and Ly$\alpha$ emission-line profiles.}
        \label{emprof_Hbg_lya}
        \hfill 
        \vspace*{0.5em}
        \centering
        \includegraphics[width=1.0\textwidth]{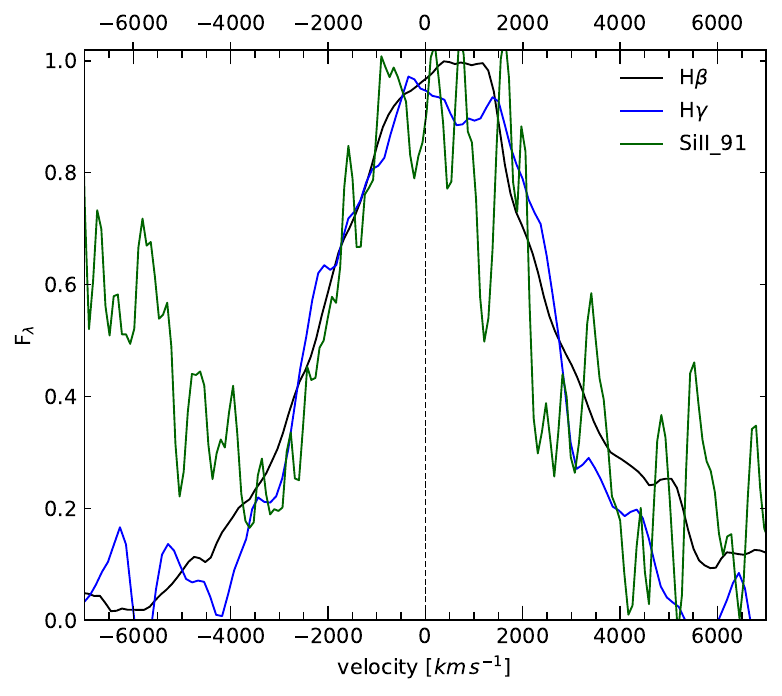}
        \caption{Scaled mean Balmer and \ion{Si}{ii}\,$\lambda$1308 (1991) emission-line profiles.}
        \label{emprof_Hbg_Si91}
        \end{minipage}
        \hfill
    % Right side with two stacked figures
        \centering
        \begin{minipage}[t]{0.46\linewidth} % First figure in right column
            \centering
            \vspace{0pt}
            \includegraphics[width=1.0\textwidth]{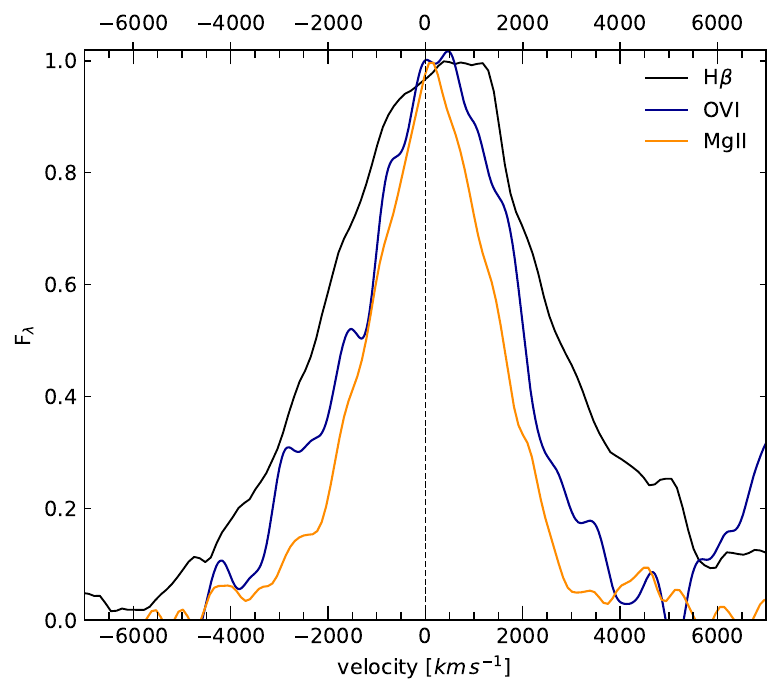}
            \caption{Scaled mean H$\beta$, \ion{O}{vi}\,$\lambda$1038, and \ion{Mg}{ii}\,$\lambda$2800 emission-line profiles.}
            \label{emprof_Hb_mg_O6}
            \hfill 
            \vspace*{0.5em}
            \centering
            \includegraphics[width=1.0\textwidth]{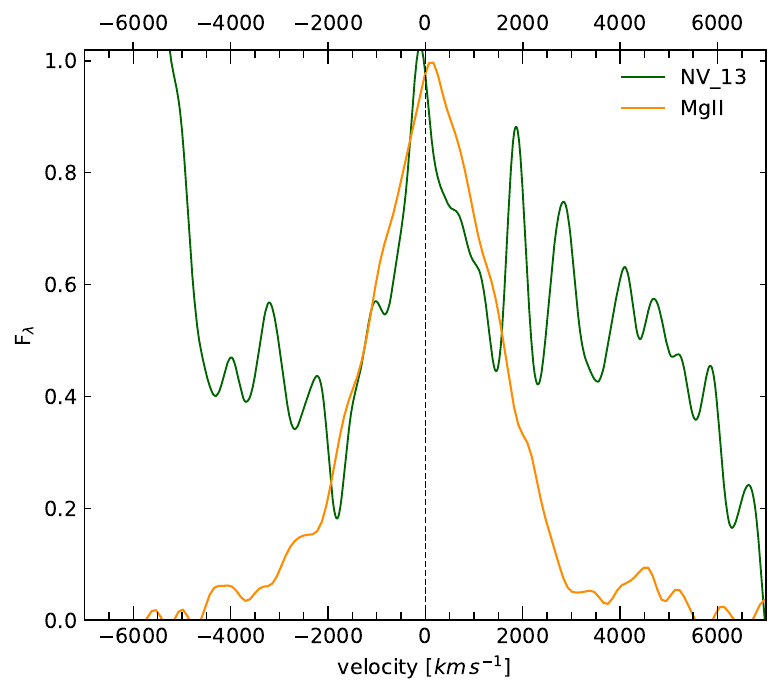}
            \caption{Scaled mean \ion{Mg}{ii}\,$\lambda$2800 and \ion{N}{v}\,$\lambda$1243 (2013) emission-line profiles.}
            \label{emprof_Mg_NV}
        \end{minipage}
\end{figure*}

Figure~\ref{emprof_Hbg_lya} shows the scaled mean profiles of
the Balmer emission lines H$\beta$ and H$\gamma$ based on the years 2013, 2019, 2022, and 2024, and Ly$\alpha$ based on the years 1991, 2013, and 2022.
\begin{table}
\tabcolsep+2.5mm
\caption{Measured line widths (FWHM) and shifts (uppermost 10\,\%) of the strongest broad emission lines.}
\begin{tabular}{llc}
\hline 
\noalign{\smallskip}
Emission Lines                &Width  &Shift \\ 
                              &[\kms] &[\kms] \\
\noalign{\smallskip}
(1)                           & (2)   & (3) \\
\noalign{\smallskip}
\hline 
\noalign{\smallskip}
\ion{O}{vi}\,$\lambda 1038$ (2013) &  3696 $\pm{}$ 200 & +309 $\pm{}$ 150 \\
Ly$\alpha$                   &  3090 $\pm{}$ 100 & -372 $\pm{}$ 100 \\
\ion{N}{v}\,$\lambda 1243$ (2013) &  2548 $\pm{}$ 400 & -107 $\pm{}$ 100 \\
\ion{O}{i}\,$\lambda 1302$+\ion{Si}{ii}\,$\lambda 1308$ (1991) &  4251 $\pm{}$ 300 & +145 $\pm{}$ 250 \\
\ion{O}{i}\,$\lambda 1302$+\ion{Si}{ii}\,$\lambda 1308$ (2013) &  5436 $\pm{}$ 400 & -687 $\pm{}$ 250 \\
\ion{Mg}{ii}\,$\lambda 2800$  &  2910 $\pm{}$ 200 & +120 $\pm{}$ 100 \\
Balmer (H$\alpha$, $\beta$, $\gamma$) & 5050 $\pm{}$ 200 & +600 $\pm{}$ 200\\
\noalign{\smallskip}
\hline
\end{tabular}
\label{em_linewidths}
\end{table}
The profiles of the Balmer lines are identical within the errors.
However, the Ly$\alpha$ emission profile is clearly narrower and its line center is blueshifted by about 1000\,\kms{} in comparison
to the Balmer lines. Blueshifted Ly$\alpha$ emission has also been 
detected before in a sample of ultraluminous galaxies (ULIRGs)  \citep{martin15}. They explained the blueshift as emission from fast, hot wind. However, our Ly$\alpha$ line
is not as broad as the Balmer lines. A certain amount of absorption in the red  Ly$\alpha$  emission wing  may have been caused by blueshifted \ion{N}{v}\,$\lambda$1243 absorption (see section \ref{sec:disc_absvar}).

The measured widths and shifts of the UV emission lines 
(Tab.\,\ref{em_linewidths}) are highly affected by strong and variable UV absorption troughs (see sections \ref{sec:disc_absvar} and \ref{sec:abslinvar}).
An identification of the $\ion{Si}{ii}\,\lambda 1306 + \ion{O}{i}\,\lambda 1303$ emission line is insecure. It seems that this line was blueshifted in 2012. 
However, it is not even ultimately 
verifiable for the year 2022. It is uncertain whether this finding is caused by emission-line variability or by highly variable absorption troughs. 
In Fig.~\ref{emprof_Hbg_Si91}, we present the mean Balmer line profiles 
together with the \ion{O}{i}\,$\lambda 1302$+\ion{Si}{ii}\,$\lambda 1308$ profile taken in 1991. They show the same profile and shift.

The width as well as the shape of the Balmer emission lines is different with respect to the
\ion{Mg}{ii}\,$\lambda$2800 line (see Fig.~\ref{emprof_Hb_mg_O6}). The Balmer lines show a FWHM of 5050.$\pm$200.
km\, s$^{-1}$, while the \ion{Mg}{ii}\,$\lambda$2800 line only has a FWHM of 
2910.$\pm$200. km\, s$^{-1}$. Furthermore, the Balmer lines show flat topped profiles. The central region of the Balmer lines is shifted by
+600.$\pm$200. km\, s$^{-1}$, while the steep \ion{Mg}{ii}\,$\lambda$2800 line is shifted by only +120.$\pm$50. km\, s$^{-1}$. 
The low-ionization \ion{Mg}{ii}\,$\lambda$2800 line shows the narrowest emission-line 
width (FWHM) of 2910.$\pm$200. \kms, 
while the other emission lines have widths of 4000. to 5000.\,\kms (see Fig.~\ref{emprof_Hb_mg_O6}).
In this plot, we show the profile of the high-ionization  \ion{O}{vi}\,$\lambda 1038$ line taken in 2013. The high-ionization \ion{O}{vi}\,$\lambda 1038$ line is broader than the low-ionization \ion{Mg}{ii}\,$\lambda$2800
line. However, both profiles (\ion{Mg}{ii}\,$\lambda$2800 and \ion{O}{vi}\,$\lambda 1038$) are narrower than those of the Balmer lines.
The \ion{O}{vi}\,$\lambda1038$ emission is clearly detected in 2013. 
However, the identification is insecure for 2022  (Fig.\ref{PG0043_UV_all_20240816}) due to its low flux and/or additional absorption.
The variable absorption troughs in that region will be discussed in more depth in sections \ref{sec:disc_absvar} and \ref{sec:abslinvar}.

Figure~\ref{emprof_Mg_NV} shows the low-ionization \ion{Mg}{ii}\,$\lambda$2800 line together with the high-ionization \ion{N}{v}\,$\lambda1243$ line. Both lines are narrow. However,
the \ion{N}{v}\,$\lambda1243$ line shows an additional strong red component at 5000 \kms{} in 2013. This red component was even stronger than the central component in 1991 (Fig.\ref{PG0043_UV_all_20240816}). It is not clear whether this red component belongs to the \ion{N}{v}\,$\lambda1243$ line or whether it has an unknown origin.

Overall it can be noted that the emission-line intensity variations and their profiles are highly peculiar. There is no clear trend that the higher ionized lines show broader profiles. There is not even any clear evidence for 
\ion{C}{iv}\,$\lambda1550$ emission.

\subsection{UV absorption-line troughs in 2022 and comparison to the years 1991 and 2013}\label{sec:disc_absvar}

\begin{figure*}[t] % Top placement
    \centering
    \hspace*{\fill}%
    % Left side with one figure
    \begin{minipage}[t]{0.46\textwidth}
        \centering
        \vspace{0pt}
        \includegraphics[width=1.0\textwidth]{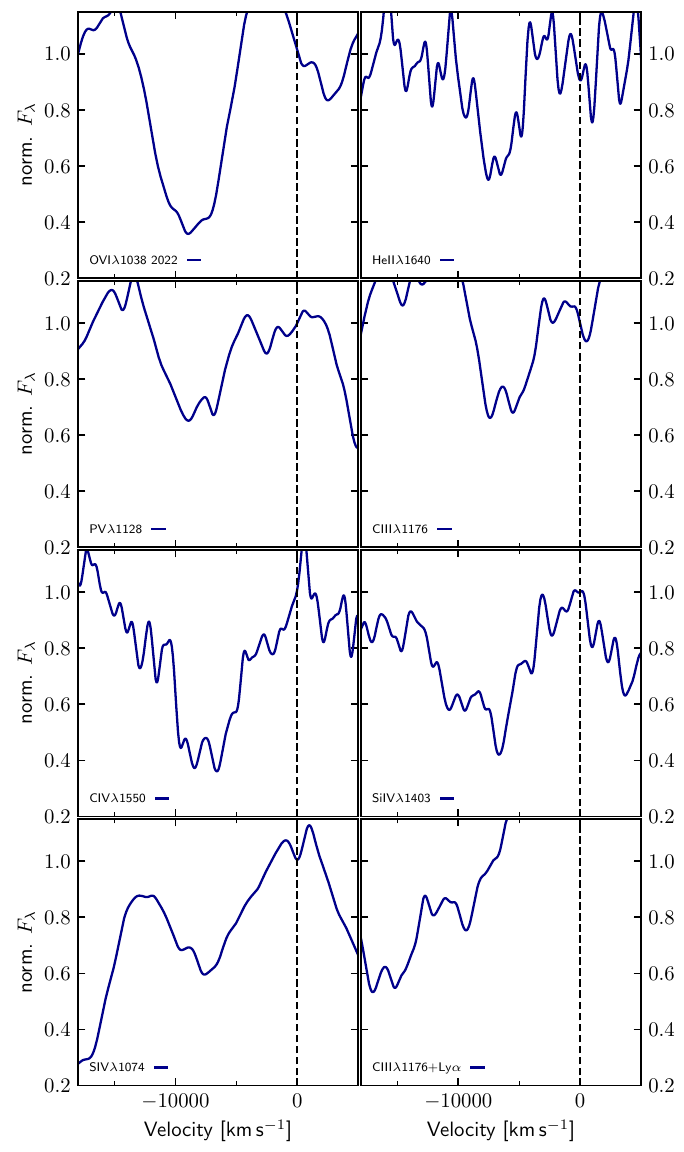}
        \caption{UV absorption-line troughs in 2022.}
        \label{profileabs_2022}
    \end{minipage}
    \hfill
    % Right side with two stacked figures
        \centering
        \begin{minipage}[t]{0.46\linewidth} % First figure in right column
            \centering
            \vspace{0pt}
            \includegraphics[width=1.0\textwidth]{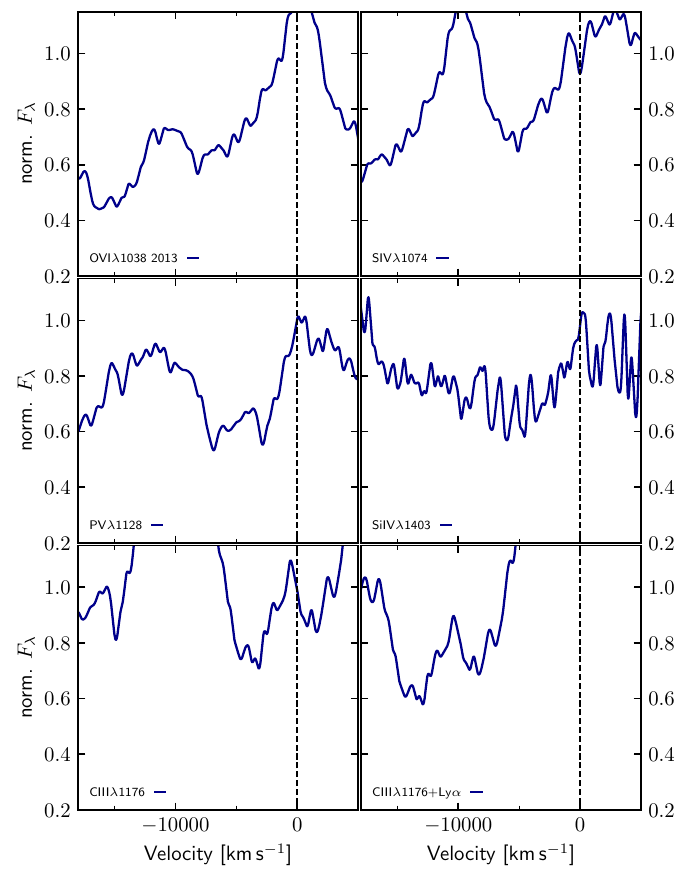}
            \caption{UV absorption-line troughs in 2013.}
            \label{profileabs_2013}
            \hfill 
            \vspace*{2em}
            \centering
            \includegraphics[width=1.0\textwidth]{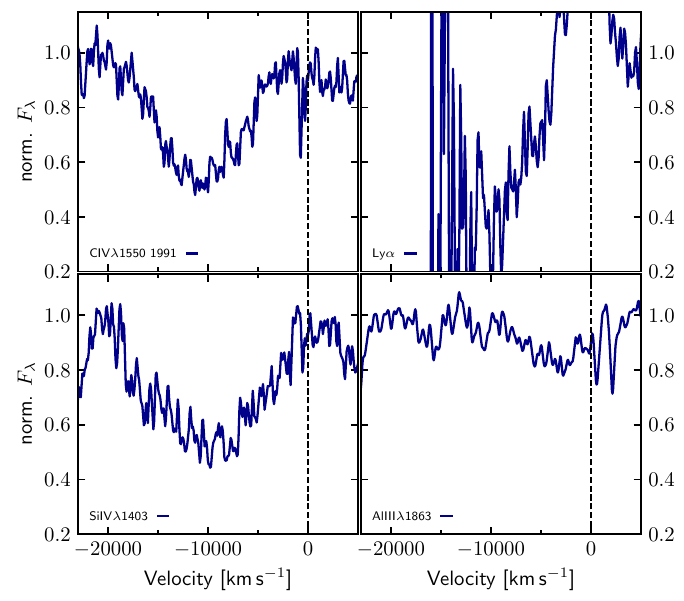}
            \caption{UV absorption-line troughs in 1991}.
            \label{profileabs_1991}
        \end{minipage}
\end{figure*}

Figure~\ref{profileabs_2022} shows the UV BAL profiles in velocity space normalized to the local continuum for the UV spectra taken in 2022. The velocities are scaled relatively to the red component of the doublet lines indicated by the dashed line. The local continua are sometimes difficult to determine because
of the overlap with additional variable emission-line components and the superposition of further absorption troughs. This applies particularly to the Ly$\alpha$ and $\ion{C}{iii}\,\lambda 1175$ lines. Therefore, the given equivalent widths of the absorptions troughs (Tab.~\ref{abs_intensities}) have errors of up to 20\,\%.  
The UV BAL profiles for the years 2013 and 1991 are shown for comparison in Fig.~\ref{profileabs_2013}  and Fig.~\ref{profileabs_1991}.

The rest equivalent widths, $W_{\lambda}$, of our measured UV absorption lines, the blue and red limits of the outflow velocities, the modified full widths at half minimum after subtracting the doublet separations, and the maximum line depths in the normalized spectra are presented in Tab.~\ref{abs_intensities}. In some cases, we cannot give a blue limit of the Ly$\alpha\,\lambda 1216$ outflow velocities 
because of the superposition of the Ly$\alpha\,\lambda 1216$ absorption trough with that of the $\ion{C}{iii}\,\lambda 1176$ trough.
The strength of a \ion{N}{v}\,$\lambda 1243$ absorption cannot been estimated because of the superposition with the variable Ly$\alpha\,\lambda 1216$ emission line and
with the Ly$\alpha\,\lambda 1216$ absorption trough. 
The following doublet separations have been taken into account
in Tab.~\ref{abs_intensities}: $\ion{O}{vi}\,\lambda 1038$: 1647 \kms; $\ion{S}{iv}\,\lambda 10741$: 2796 \kms; $\ion{P}{v}\,\lambda 1128$: 2660 \kms; $\ion{Si}{iv}\,\lambda 1403$: 1924 \kms; $\ion{C}{iv}\,\lambda 1550$:  503 \kms;  $\ion{Al}{iii}\,\lambda 1863$: 1288 \kms.
\begin{table*}
%\centering
%\tabcolsep+3.5mm
\tabcolsep+1.mm
\caption{UV absorption trough equivalent widths, outflow velocities (max/min), FWHM, and their depths.}
\begin{tabular}{l|rcrc|rcrc|rcrc}
\hline 
\rule{0pt}{2.5ex}            & \multicolumn{4}{c}{1991} & \multicolumn{4}{c}{2013} & \multicolumn{4}{c}{2022} \\
\hline
%\noalign{\smallskip}
Absorption &$W_{\lambda}$  & $v_{outfl.}$ & FWHM & Depth &$W_{\lambda}$  & $v_{outfl.}$ & FWHM & Depth & $W_{\lambda}$  & $v_{outfl.}$ & FWHM & Depth \rule{0pt}{2.5ex} \\ 
       &[\AA{}]  & [\kms] & [\kms] &  & [\AA{}]  & [\kms] & [\kms] & & [\AA{}]  & [\kms] & [\kms] & \\
%\noalign{\smallskip}                                                                      
(1)      & (2)  &  (3) & (4) &(5) &(6) &(7) &(8) &(9) &(10) &(11) &(12) &(13) \rule{0pt}{2.5ex} \rule[-1.2ex]{0pt}{0pt} \\
%\noalign{\smallskip}
\hline 
%\noalign{\smallskip}
$\ion{O}{vi}\,\lambda 1038$ & & & & &4.6:&-9620/-1150&4710 &0.43  &12.9 &-13450/-4750& 4130&0.64  \rule{0pt}{2.5ex} \\
$\ion{S}{iv}\,\lambda 1074$ &  & & & &5.5&-8320/-1300&2420 &0.35 & 6.8& -11910/-2110& 3120&0.41 \\
$\ion{P}{v}\,\lambda 1128$   &  & & & & 10.6&-10950/-50& 4360& 0.47 & 7.0& -12200/-4540& 2300& 0.35\\
$\ion{C}{iii}\,\lambda 1176$  & & & & & 3.3& -5880/-890& 2930:& 0.29 & 5.4& -9010/-3380 & 4030& 0.34\\
Ly$\alpha\,\lambda 1216$                  &  & /-3670& &0.65:& &-16340:/-6130& 3430:& 0.32:  & & /-7240 & & 0.19:\\
$\ion{Si}{iv}\,\lambda 1403$ &25.3&-18710/-1220& 11980&0.56&18.9&-17230/0& 12110&0.43& 15.2& -11940/-420& 4290& 0.58 \\
$\ion{C}{iv}\,\lambda 1550$ &21.9& -19660/-1030& 9210&0.51& & & & & 24.6& -16120/-50& 4920 & 0.64 \\
$\ion{He}{ii}\,\lambda 1640$  &  & & &     & & & & & 9.9 & -10260/-4430& 3720& 0.45\\
$\ion{Al}{iii}\,\lambda 1863$ & 6.7& -8750/3000& 4640& 0.22     & & & &       & & & & \rule[-1.2ex]{0pt}{0pt} \\
%\noalign{\smallskip}
\hline
\end{tabular}\\
\tablefoot{Given are the modified full widths at half minimum after subtracting the doublet separations.  Errors for the equivalent widths reach up to 20\%, and 5 to 10\% for the FWHM and line depths. The outflow velocities have an error of 5\%. The errors are larger in some cases due to unclear continuum intensities as well as superpositions of absorption troughs. They are marked by the symbol : .}
\label{abs_intensities}
\end{table*}

PG\,043+039 is an extreme LoBAL based on i) its
very strong \ion{C}{iv} BAL absorption, ii) its unusually strong $\ion{P}{v}$ absorption, 
as well as iii) its large velocity shift of nearly -20,000 \kms in the 
\ion{C}{iv} BAL. 
PG\,043+039 exhibits $\ion{P}{v}\,\lambda 1128$ equivalent widths of 7--10 \AA{}.
Such high values are typical for $\ion{P}{v}$ BALs \citep{hamann19}.
The resolved $\ion{P}{v}\,\lambda 1128$ doublet shows an observed saturated depth ratio near 1:1. This is a typical ratio, as seen elsewhere in other BALs (\citealt{hamann19}, and references therein).
Such a ratio requires optical densities of log $N_{H}$ (cm$^{-3}$)  $\gtrsim22.7 $ 
and large ionization parameters of log U $\gtrsim$ -0.5 \citep{hamann19}.  It implies that most UV outflow line ratios such as $\ion{Si}{iv}\,\lambda1394,1403$ or  $\ion{O}{vi}\,\lambda1032, 1038$ are also saturated.
A detection of the low abundance-line $\ion{P}{v}\,\lambda 1128$ doublet implies large column densities. The \ion{C}{iv} trough does not reach zero intensity. This indicates that the outflowing gas only partially covers the UV continuum source by a clumpy inhomogeneous medium (see also \citealt{capellupo17}). 

We now compare the absorption troughs among each other for the individual observations in the years 2022, 1991, and 2013, as well as in comparison to each other. We do this because the general absorption shifts are different for the individual years.
The highly ionized absorption lines such as
$\ion{O}{vi}\,\lambda 1038$, 
$\ion{C}{iv}\,\lambda 1550$, $\ion{Si}{iv}\,\lambda 1403$, 
$\ion{S}{iv}\,\lambda 1074$, and $\ion{P}{v}\,\lambda 1128$ show the largest velocity shifts in their BALs. 
The flux minima in lower ionized BALs such as $\ion{C}{iii}\,\lambda 1176$ and 
$\ion{Al}{iii}\,\lambda 1863$ have smaller blueshifts (see Figs. \ref{profileabs_2022}, \ref{profileabs_2013}, and \ref{profileabs_1991}, and 
Tab.~\ref{abs_intensities}).
This is consistent with detailed studies on other QSO BAL outflows
\citep{hamann19}.
Weak $\ion{He}{ii}\,\lambda 1640$ emission-line intensity is correlated with larger BAL outflow speeds \citep{hamann19}. This is consistent with
the spectrum taken in 1991 where we see the largest outflow velocities. There is no clear  $\ion{He}{ii}\,\lambda 1640$ emission line detection for the year 2022 when the outflow velocities were lower.

In addition, we found strong evidence for the presence of
$\ion{He}{ii}\,\lambda 1640$ absorption in the 2022 spectrum (see Figs. \ref{PG0043_UV_22_20240813} and \ref{profileabs_2022}). This absorption has not been identified in the $\ion{P}{v}$ BAL samples of \cite{capellupo17} and \cite{hamann19}.
Furthermore, the spectra in their samples always show 
weaker Ly$\alpha$ emission relative to
\ion{N}{v}\,$\lambda1243$ emission or no Ly$\alpha$ emission at all. In contrast
to these samples, PG\,0043+039 exhibits strong Ly$\alpha$ emission, which is typical for normal quasars.

\subsection{Interpretation of the combined absorption- and emission-line variations in the far-UV} \label{sec:optFUVcombvar}
We present a possible interpretation of the far-UV spectra
of PG\,0043+039 in 2013 \citep{kollatschny15} . The UV spectrum (shortward of 1600 \AA) showed very bumpy structures in addition to only a few clear emission lines such as Ly$\alpha$.
It was not clear whether the bumpy structure is caused by strong absorption features or by strong emission components. In \cite{kollatschny15}, we proposed cyclotron emission lines as the cause for the observed broad bumps -- based on modeling of the far-UV spectra.

These emission bumps as well as the absorption structures did not remain constant in intensity and wavelength space in the 2022 data. In principle, we can reproduce 
the blueshifted spectral distribution of UV emission lines (or absorption troughs) for the year 2022 based on cyclotron lines for slightly higher field strengths than in 2013. An increase of the magnetic field leads to simultaneous wavelength shifts of all  
cyclotron lines and their higher harmonics \citep{kollatschny15}.
However, based on the new spectra taken in 2022 and in comparison with other extreme $\ion{P}{v}$ BAL sources (e.g., \citealt{borguet13}, \citealt{capellupo17}, \citealt{hamann19}), it now seems more probable that the observed UV spectral structures are caused by broad absorption troughs 
(see Sect.~\ref{sec:optUVcontvar} and \ref{sec:disc_absvar}). On the other hand, it is still difficult to explain the emission component redward of the $\ion{N}{v}\,\lambda 1243$ line (see Figs.~\ref{PG0043_UV_22_20240813}, \ref{emprof_Mg_NV}) in an easy way.

\subsection{General absorption-trough variations in velocity space for the different epochs} \label{sec:abslinvar}

We show the HST UV spectra for the years 1991, 2013, and 2022 in Fig.~\ref{PG0043_UV_all_20240816}. There, you can see clear variations in the absorption-line troughs.
Furthermore, we present UV difference spectra in
Fig.~\ref{PG0043_UV_diffs_20250620}. One can recognize that there are strong variations in the absorption troughs
as well as in the emission lines.
To study the absorption-line variability in more detail, in 
Figs. \ref{profileabs_2022}, \ref{profileabs_2013}, \ref{profileabs_1991}, and 
Tab.~\ref{abs_intensities} we show 
the UV BAL profiles in velocity space normalized to the local continuum for the years 2022, 2013, and 1991. Besides the general trend that highly ionized absorption lines show the largest blueshifts and less ionized
lines show smaller blueshifts, there is superimposed a second trend: all the absorption troughs of 2013 are redshifted with respect to the troughs of 2022 and all the absorption troughs of 1991 are blueshifted with respect to the troughs of 2022.

\begin{figure*}
\sidecaption
        \includegraphics[width=12 cm, angle=0] {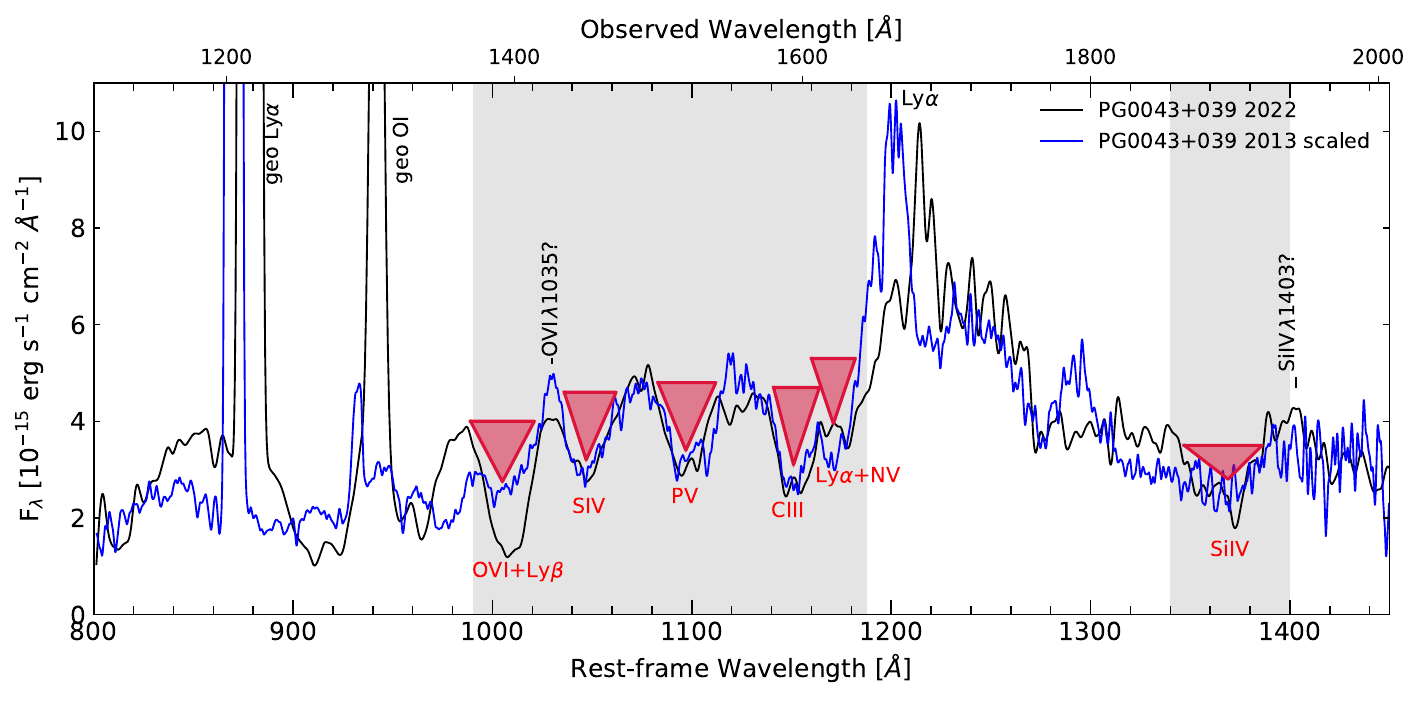}
        \caption{UV absorption troughs of spectrum taken in 2022 and scaled 2013 spectrum.
        The 2013 spectrum has been multiplied by a factor of 0.692 and blueshifted by 2740 \kms. The troughs of the individual absorption lines are indicated by red triangles. The shaded areas give the wavelength regions that were used
        for cross-correlating the wavelength shifts of the two spectra.}
        \label{uvabsvar2022_2013_velcor1}
\end{figure*}

\begin{figure*}
\sidecaption
        \includegraphics[width=12 cm, angle=0] {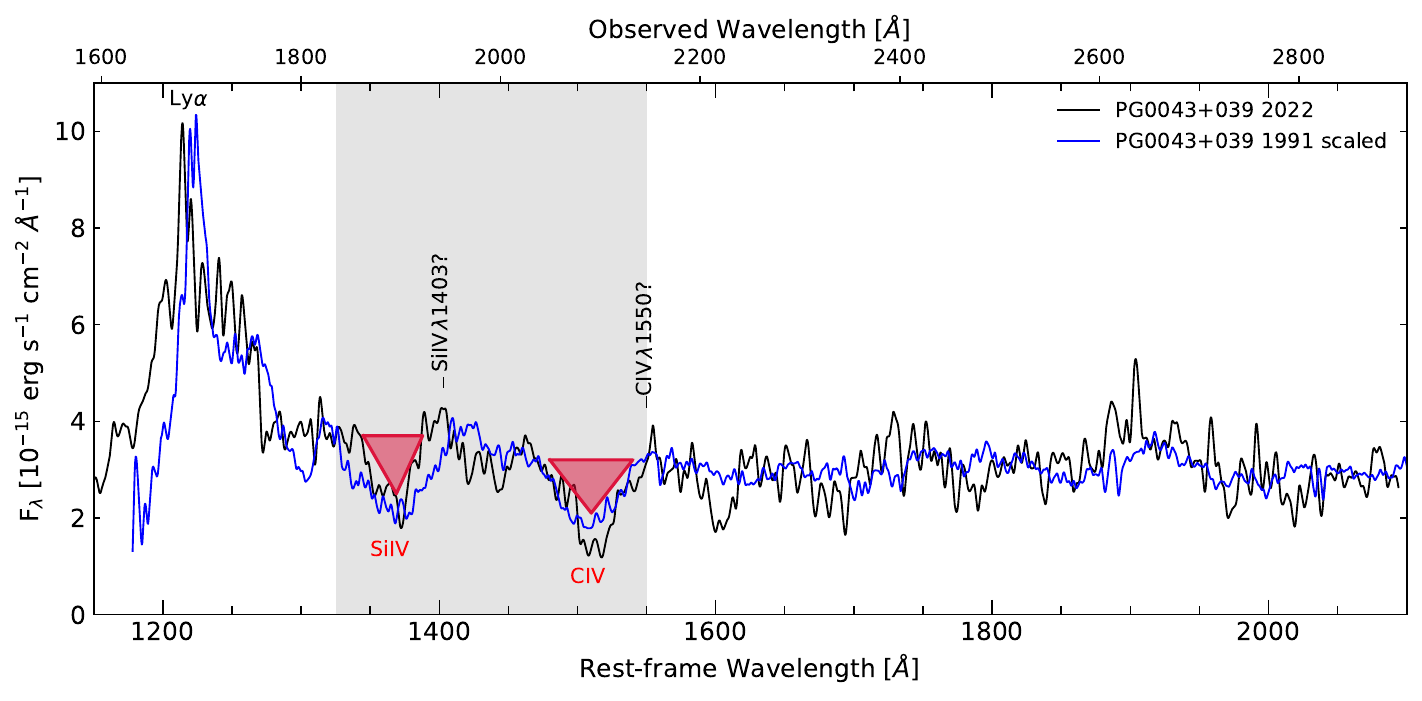}
        \caption{UV absorption troughs of spectrum taken in 2022 and  scaled 1991 spectrum.
        The 1991 spectrum has been multiplied by a factor 0.74 and redshifted by 2000 \kms. The troughs of the  $\ion{Si}{iv}\,\lambda 1403$ and 
        $\ion{C}{iv}\,\lambda 1550$ absorption lines are indicated by red triangles. The shaded area gives the wavelength region that was used
        for cross-correlating the wavelength shifts of the two spectra.}
        \label{uvabsvar2022_1991_velcor2}
\end{figure*}

To compare the variability of the general absorption pattern, in  Fig.~\ref{uvabsvar2022_2013_velcor1} we show the scaled 2013 spectrum with respect to that of 2022. The 2013 spectrum has been multiplied by a factor of 0.692 and blueshifted by 10.5 \AA{} corresponding to 
2740.$\pm{}$200.\,\kms.
Similarly, in Fig.~\ref{uvabsvar2022_1991_velcor2} we compare the 
general absorption pattern of the scaled 1991 spectrum to that of 2022.
The 1991 spectrum was multiplied by a factor 0.74 and 
redshifted by 9.5 \AA{} corresponding to 2000.$\pm{}$300.\,\kms.
\begin{figure*}[t] % Top placement
    \centering
    \hspace*{\fill}%
    % Left side with one figure
    \begin{minipage}[t]{0.46\textwidth}
        \centering
        \vspace{0pt}
        \includegraphics[width=1.0\textwidth]{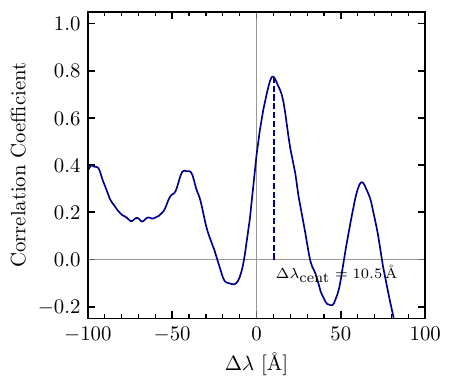}
        \caption{CCF deriving wavelength shift of absorption-trough regions (see Fig.~\ref{uvabsvar2022_2013_velcor1}) of the 2013 spectrum with respect to 2022 spectrum.}
        \label{CCF2013to2022}
    \end{minipage}
    \hfill
    % Right side 
        \centering
        \begin{minipage}[t]{0.46\linewidth} % First figure in right column
            \centering
            \vspace{0pt}
            \includegraphics[width=1.0\textwidth]{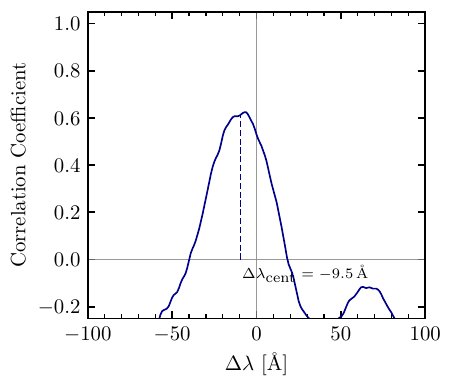}
            \caption{CCF deriving wavelength shift of absorption-trough region (see Fig.~\ref{uvabsvar2022_1991_velcor2}) of 1991 spectrum with respect to 2022 spectrum.}
            \label{CCF1991to2022}
        \end{minipage}
\end{figure*}
We derived these $\lambda$-shifts by cross-correlating the spectra in the regions where the strong absorption troughs of 
$\ion{O}{vi}\,\lambda 1038$, $\ion{S}{iv}\,\lambda 1074$, 
$\ion{P}{v}\,\lambda 1128$,  $\ion{C}{iii}\,\lambda 1176$, and 
Ly$\alpha\,\lambda 1216$ are present (990 -- 1190\AA{}), as well as 
$\ion{Si}{iv}\,\lambda 1403$ (1340 - 1400\,\AA{}), for the years 2022 and 2013 (see the shaded regions in Fig.~\ref{uvabsvar2022_2013_velcor1}).      
Furthermore, we correlated the wavelength region that contains the absorption troughs of $\ion{Si}{iv}\,\lambda 1403$ and
$\ion{C}{iv}\,\lambda 1550$ (1325 to 1550\,\AA{}; see the shaded regions in Fig.~\ref{uvabsvar2022_1991_velcor2})
to derive the shift of the 1991 spectrum with respect to the 2022 spectrum.
We used the interpolation cross-correlation function (CCF) method 
developed by \cite{gaskell87} to calculate the
$\lambda$-shifts. We generated our own CCF code (\citealt{dietrich95}, \citealt{kollatschny18})
based on similar assumptions.  The derived CCFs are presented in Figs~\ref{CCF2013to2022} and \ref{CCF1991to2022}.
The CCFs show clear and unique maxima.
We determined the centroids of these CCF using
only the parts of the CCFs above 80\% of the peak value,
$r_{max}$. A threshold value of 0.8 $r_{max}$ is generally a good choice,
as has been shown by, for example, \cite{peterson04}. We
derived the uncertainties in our cross-correlation results based on the widths at the top of the CCFs. 

All the strong absorption-line troughs varied in unison in velocity space back and forward without any major changes in their absorption strengths or profiles.
It is remarkable that the profiles and equivalent widths of all the absorption troughs remained so similar - despite different onset velocities
-- for the years 2013 and 2022. The 2013 spectrum is blueshifted by 
2740\,\kms{} in Fig.~\ref{uvabsvar2022_2013_velcor1}. The same trend is true when we compare the profiles and equivalent widths for the years 1991 and 2022
in Fig.~\ref{uvabsvar2022_1991_velcor2}. However, in this case the 1991 spectrum is redshifted by 2000\,\kms. Here, we see additional weak variability in the shapes of the absorption profiles.
In Tab.~\ref{cont_absvel}, we present the UV-continuum fluxes at 1288\,\AA{}
for the years 1991, 2013, and 2022, and the general outflow velocities of the BAL troughs with respect to the year 2022.
\begin{table}
%\centering
%\tabcolsep+3.5mm
\tabcolsep+5.5mm
\caption{Relation between UV-continuum flux at 1288\,\AA{} and relative BAL outflow velocity.}
\begin{tabular}{lcc}
\hline 
\noalign{\smallskip}
Year              & continuum flux      &shift \\ 
  &  [10$^{-15}$\,erg\,s$^{-1}$\,cm$^{-2}$\,\AA$^{-1}$] &[\kms] \\
\noalign{\smallskip}
(1)                           & (2)   & (3) \\
\noalign{\smallskip}
\hline 
\noalign{\smallskip}
1991  &  7.60 $\pm{}$ .3 & -2000 $\pm{}$ 300 \\
2013  &  8.03 $\pm{}$ .3 & +2740 $\pm{}$ 200 \\
2022  &  6.24 $\pm{}$ .3 & 0 \\
\noalign{\smallskip}
\hline
\end{tabular}
\label{cont_absvel}
\end{table}
It is immediately clear that there is no correlation between UV-continuum luminosity and the outflow velocity of the BAL troughs.

One possible explanation for the unison redshift and subsequent blueshift of all the absorption troughs might be caused by the fact that the wind leaving the disk shares the disk rotation.
The timescale of such a revolution scenario of the accretion disk is on the order of years (see, e.g., \citealt{gibson08a}). Therefore, we might see different wind shares at different epochs.

\subsection{Correlations between continuum variations and BAL variations}\label{sec:correlation}

The question remains as to whether the observed variations in the continuum flux are caused by varying absorption of the central ionizing source. However, there is no indication of dust absorption based on X-ray-continuum fitting. The X-ray spectrum -- taken in 2013 -- was compatible with a normal quasar power-law spectrum with moderate intrinsic absorption \citep{kollatschny16}. The measured low Balmer decrement also indicates low dust absorption.

It is possible that the weak X-ray flux in PG\,0043+039
is intrinsic or caused by (partial) covering of the central ionizing source.
\cite{saez21} found indications of an X-ray-wind connection in PG\,2112+059.
They observed a correlation between an increase of the BAL equivalent widths with a significant dimming in soft X-ray emission 
($\lessapprox$2keV). They claim that this is in agreement with increased absorption.
They suggest a wind-shield scenario where the outflow inclination with respect to the line of sight is decreasing and/or the wind mass is increasing. 
BAL profiles are largely determined by the 
velocity-dependent geometric covering factor of saturated absorbers \citep{arav99}.
\cite{green23} investigated the origin of the absorption-line variability in the nearby Seyfert galaxy WPVS 007 over seven years. They diagnosed the primary cause of the
absorption-line variability in WPVS 007 to be a change in the covering fraction of the continuum by the outflow. 
\cite{giustini23} carried out a coordinated study of UV absorption troughs together with X-ray observations of PG\,1126-042. They claim that the absorption from the
X-ray partially covering gas and from the blueshifted $\ion{C}{iv}$ troughs appear to vary in a coordinated manner.

Our study shows that the UV-continuum flux decreased in PG\,0043+039 by a factor of more than 1.3 between the years 2013 and 2022 (see Fig.~\ref{PG0043_UV_all_20240816}) and the X-ray flux decreased by a factor of 3.4. 
However, the equivalent widths of the UV absorption troughs (see Fig.~\ref{uvabsvar2022_2013_velcor1}) varied only marginally after correction for the velocity shift. 
At first approximation we do not see a general indication for a correlation between the variation of the photoionizing continuum and the absorber variation -- as it is also being discussed in QSO BAL variability studies on multiyear timescales \citep{gibson08a}.
However, based on the simultaneous UV and X-ray observations - taken in 2013 and 2022 - we see higher maximum velocities of the blueshifted broad absorption lines in the UV when the X-ray flux was lower. 
\cite{proga05} suggested that UV-line-driven winds may affect the X-ray corona.
\cite{gallagher06} found a likely correlation between the maximum outflow velocity of $\ion{C}{iv}$ absorption and the magnitude of X-ray
weakness as well.
Our finding is consistent with results from 
\cite{zappacosta20}, which reported an anticorrelation between the X-ray luminosities and the blueshifted velocities of the \ion{C}{iv} emission line in a sample of luminous quasars.
They interpret their results in the context of accretion disc winds
that are reducing the coronal X-ray production.

\section{Summary}\label{sec:summary}

We present combined (optical, UV, X-ray) spectral and photometric variability results of the most X-ray-faint AGN PG\,0043+039 -- carried out with the HST, the HET, and XMM-Newton. 
 Our findings can be summarized as follows:

\begin{enumerate}[(1)]

\item PG\,0043+039 exhibited an extremely low X-ray luminosity and showed an extreme steep $\alpha_{ox}$ value of $-$2.47 in 2022. These values are similar to those measured in 2005. The X-ray luminosity was a factor of 3.4 higher in the meantime in 2013. The value of $\alpha_{ox}$ was  
$-$2.37 at that time.

\item The optical and UV continuum decreased by only a factor of 1.3 -- 1.5 
    from 2013 to 2022.

\item The optical-near-UV continuum could be reproduced with a
simple power law of $f_\lambda \sim\lambda^{-\alpha}$ with a gradient $\alpha$=1.317 in 2022. 
The gradient was even stronger with a value of $\alpha$=1.69  when PG\,0043+039 was in a high state in 2013. 
The observed continuum gradient changes to an inverted value in the far-UV at a turning point of 2400\,\AA{} at all times from 1991 till 2022.

\item Very strong emission line intensity variations  
by factors of eight or more were observed in the
$\ion{O}{vi}\,\lambda 1038$ and $\ion{Si}{iv}\,\lambda 1403$ lines between 2013 and 2022. The other UV emission lines such as Ly$\alpha$  only decreased by a factor of 1.4. These values are affected by major errors because of additional variations of the UV absorption troughs. The optical emission lines showed only minor intensity variations in contrast to the UV lines.

\item PG\,0043+039 exhibits strong and broad absorption lines in the UV. 
 The highly ionized absorption lines show the largest velocity blueshifts in their BALs. 
Lower ionized BALs have smaller blueshifts. 

\item The observed $\ion{P}{v}$ absorption is very strong, pointing to large column densities. On the other hand, the $\ion{C}{iv}$ trough does not reach zero intensity. This indicates that the outflowing gas only partially covers the central ionizing source. PG\,0043+039 can be classified as one of the rare $\ion{P}{v}$ BALQs. 

\item PG\,0043+039 shows strong Ly$\alpha$ emission despite strong $\ion{P}{v}$ absorption. PG\,0043+039 is the only source with strong Ly$\alpha$ in the sample of strong $\ion{P}{v}$ absorbers. 

\item All the strong UV absorption-line troughs varied in unison in velocity space back and forward (-2000. $\pm{}$ 300. \kms, +2740. $\pm{}$ 200. \kms) without any major changes in absorption strength as well as in their profiles for the years 1991, 2013, and 2022.

\item We found no general correlations of the X-ray/opt/UV continuum variations with the broad absorption-line variations. However, based
on the simultaneous UV and X-ray observations -- taken in
2013 and 2022 -- we see higher maximum velocities of the
blueshifted broad absorption lines in the UV when the X-ray
flux was lower.

\end{enumerate}

\begin{acknowledgements}
This work has been supported by the DFG grants KO857/35-1 and KO857/35-2. This paper is based on observations obtained with the Hobby-Eberly Telescope, which is a joint project of the University of Texas at Austin, the Pennsylvania State University,  Ludwig-Maximilians-Universität M\"unchen, and Georg-August-Universit\"at G\"ottingen. Furthermore, it is based on observations obtained with 
the Hubble Space Telescope and XMM-Newton, an ESA science mission with instruments and contributions directly funded by ESA Member States and NASA. This research has made use of data from the NuSTAR mission, a project led by the California Institute of Technology, managed by the Jet Propulsion Laboratory, and funded by the National Aeronautics and Space Administration. Data analysis was performed using the NuSTAR Data Analysis Software (NuSTARDAS), jointly developed by the ASI Science Data Center (SSDC, Italy) and the California Institute of Technology (USA).
\end{acknowledgements}

\bibliographystyle{aa} % style aa.bst
\bibliography{literature} % your references Yourfile.bib

%%%%%%%%%%%%%%%%%%%%%%%%%%%%%%%%%%%%%%%%%%%%%%%%%%%%%%%%%%%%%%%%%%%%%%%%%%%%5
\end{document}